\documentclass[preprint,preprintnumbers,amsmath,amssymb,10pt]{revtex4}
\usepackage[colorlinks,linkcolor=red,citecolor=blue]{hyperref}
\usepackage{epsf,epsfig}
\usepackage{bm}
\usepackage{graphicx}
\usepackage{natbib}
\usepackage{color}
\usepackage{mathrsfs}
\usepackage{float}
\usepackage{multirow}
\usepackage{makecell}
\usepackage{amssymb}
\usepackage{amsmath}
\usepackage{xcolor}
\usepackage{url}
\usepackage{adjustbox}
\usepackage{booktabs}

\newcommand{\commentold}[1]{}
\DeclareMathSymbol{:}{\mathpunct}{operators}{"3A}
\begin{document}

\title{\textbf{Alleviating the $H_0$ tension in Rastall gravity}}

\author{R. Mohebi\footnote{rozita.mohebi@uok.ac.ir}, Kh. Saaidi\footnote{ksaaidi@uok.ac.ir}, T. Golanbari\footnote{t.golanbary@uok.ac.ir} and K. Karami\footnote{kkarami@uok.ac.ir}}
\affiliation{\small{Department of Physics, University of Kurdistan, P.O. Box 66177-15175, Sanandaj, Iran}}

\date{\today}

\begin{abstract}
	The persistent discrepancy between local determinations of the Hubble constant $H_0$ and the Planck 2018 value ($67.4 \pm 0.5~{\rm km\,s^{-1}\,Mpc^{-1}}$) within $\Lambda$CDM remains a central challenge in precision cosmology. We investigate the Hubble tension in $\Lambda$CDM and its Rastall extension (R-$\Lambda$CDM) for flat, open, and closed geometries.
	
	We analyze three primary dataset combinations: D$_1$ (late-time probes: SN + $H(z)$ + $f\sigma_8$), D$_2$ (late-time probes combined with DESI DR2 BAO and BBN), and D$_3$ (late-time probes combined with BAO and Planck 2018 CMB distance priors). Parameters are constrained via Markov Chain Monte Carlo sampling, and tensions with SH0ES ($73.2 \pm 1.3~{\rm km\,s^{-1}\,Mpc^{-1}}$) and Planck are expressed in units of the combined uncertainty.
	In addition, we include a Planck-only configuration (D$_4$) as a reference baseline to isolate early-Universe constraints on $H_0$.
	Within $\Lambda$CDM, D$_1$ and D$_2$ yield $H_0 \simeq 70.75$-$71.43~{\rm km\,s^{-1}\,Mpc^{-1}}$, reducing the SH0ES discrepancy to $1.11\sigma$--$1.63\sigma$ while maintaining a $3.62\sigma$-$4.23\sigma$ tension with Planck. Including CMB distance priors (D$_3$) shifts the result to $H_0 \simeq 67.18$--$67.55~{\rm km\,s^{-1}\,Mpc^{-1}}$, consistent with Planck at $0.09\sigma$-$0.38\sigma$ but increasing the SH0ES discrepancy to $4.25\sigma$-$4.51\sigma$.
	
	A similar dataset-driven behavior is found in R-$\Lambda$CDM. For D$_1$ and D$_2$, we find $H_0 \simeq 70.79$--$71.48~{\rm km\,s^{-1}\,Mpc^{-1}}$, with SH0ES tensions of $1.11\sigma$-$1.60\sigma$ and Planck discrepancies of $3.64\sigma$-$4.19\sigma$. For D$_3$, the inferred value becomes $H_0 \simeq 68.50$--$69.50~{\rm km\,s^{-1}\,Mpc^{-1}}$, reducing the Planck tension to $1.70\sigma$-$2.95\sigma$ and yielding a SH0ES discrepancy of $2.65\sigma$-$3.45\sigma$. In this configuration, $H_0$ lies between the Planck and SH0ES determinations, partially alleviating their discrepancy. This behavior can be interpreted as a consequence of the modified matter evolution in Rastall cosmology, which induces a degeneracy between $\epsilon$ and $H_0$ at the background level. Small positive values $\epsilon \sim \mathcal{O}(10^{-3})$, favored by CMB-inclusive datasets, lead to a mild but systematic increase in $H_0$ relative to $\Lambda$CDM, while the effect vanishes as $\epsilon \to 0$.
	
	Model comparison using AIC and BIC shows statistical equivalence between $\Lambda$CDM and R-$\Lambda$CDM for D$_1$ and D$_2$, while in D$_3$ the R-$\Lambda$CDM extension is preferred in flat and closed geometries and decisively favored in the open case. 
	Overall, the results highlight that any apparent alleviation of the Hubble tension remains strongly dependent on dataset composition and does not provide a universal resolution within the considered framework.
	
	\textbf{keyword:} Hubble tension, Rastall gravity 
\end{abstract}

\maketitle

\section{Introduction}\label{Introduction}

Modern cosmology is currently confronted with several significant challenges, among which the Hubble constant $H_0$ tension and the nature of dark energy stand out as particularly prominent. Understanding these issues may point toward new physics and has therefore motivated extensive theoretical and observational investigations. Various experiments have been conducted to explore the properties of dark energy and the accelerated expansion of the Universe. Cosmologists have been striving to find an appropriate model to explain the late-time positive cosmic acceleration, leading to extensive research on gravitational theories \cite{riess1998observational, perlmutter1999measurements, spergel2003first, peebles2003cosmological, Pavlov:2013nra, Adil_2023, Escamilla_2024, Vincenzi_2024}.

Einstein’s General Relativity (GR), supplemented by a cosmological constant, remains the preferred theory of gravity for explaining the accelerated expansion of the Universe. However, GR encounters conceptual and phenomenological challenges on cosmological scales, including singularities and parameter degeneracies. These limitations have stimulated extensive efforts toward extending or modifying the gravitational framework. Inflationary cosmology is one of the most successful theoretical developments in modern cosmology. Although originally proposed to address the classical problems of the Big Bang scenario, its predictions, such as the generation of primordial quantum fluctuations, and its remarkable agreement with observations have established it as a cornerstone of contemporary cosmological theory \cite{Starobinsky:1980te,Guth:1980zm, Albrecht:1982wi, Linde:1981mu, Linde:1983gd, Spalinski:2007dv, Bessada:2009pe, Nazavari:2016yaa, Amani:2018ueu, Mohammadi:2018zkf, berera1995warm, berera2000warm, BasteroGil:2004tg, Rosa:2018iff, Bastero-Gil:2019gao, Sayar:2017pam, Akhtari:2017mxc, Sheikhahmadi:2019gzs, Bastero-Gil:2017wwl, Motohashi:2014ppa, Odintsov:2017hbk, Oikonomou:2017bjx, Mohammadi:2019qeu, Mohammadi:2020ftb, Mohammadi:2022vru, Yogesh:2025wak, Mohammadi:2025gbu}. Although inflation pertains to the earliest moments of cosmic history, its consequences extend to late-time observables, including implications for the current Hubble rate. 

To explain the observed cosmic acceleration, the cosmological constant and dark energy remain popular candidates. Dark energy models constitute one of the most well-known approaches to describe the late-time acceleration of the Universe. On the other hand, cosmologists have explored various modified gravity theories, such as $f(R)$ gravity \cite{DeFelice:2010aj, Sotiriou:2008rp, nojiri2011unified}, $f(T)$ gravity \cite{Bengochea:2008gz, Wu:2010mn, Myrzakulov:2010vz}, $f(R, T)$ gravity \cite{harko2011f, Carvalho:2017pgk, Mohammadi:2023kzd, Ossoulian:2023moq, Taghavi:2023ptn}, modified Gauss-Bonnet gravity $f(G)$ \cite{Li:2007jm, Davis:2007ta, Yogesh:2025hll}, scalar-tensor gravity \cite{Brans:1961sx, Carloni:2007eu, Mohammadi:2022fiv}, brane gravity \cite{Maartens:2010ar, Borzou:2009gn, Mohammadi:2020ake}, among others. Modified gravitational theories generally extend the Einstein--Hilbert action of General Relativity through modifications of the curvature-based description of gravity. In this context, Rastall gravity has emerged as a phenomenologically interesting alternative, based on a generalized conservation law for the energy--momentum tensor \cite{AFSHAR2023101357, SALTI2020100630, SALEEM2021100808, Nazavari:2023khn}.

Rastall gravity was originally proposed as a modification of the standard conservation law of the energy--momentum tensor \cite{Rastall1972, Rastall1976}. In contrast to General Relativity, where the covariant divergence of the energy--momentum tensor vanishes, Rastall theory allows a non-zero divergence proportional to the gradient of the Ricci scalar, leading to an effective non-minimal coupling between matter and geometry.

The theoretical status of Rastall gravity has been the subject of ongoing discussion. In particular, it has been argued that the theory can be formally recast as standard General Relativity with a redefined (non-conserved) effective energy--momentum tensor \cite{Visser2017}. On the other hand, several studies have pointed out that, depending on the physical interpretation and the matter sector, the theory may still lead to phenomenological deviations from GR in specific astrophysical and cosmological contexts \cite{Darabi2018}. In this work, we adopt the commonly used phenomenological viewpoint and treat Rastall gravity as an effective framework to test possible deviations from the standard conservation law at cosmological scales.

From an observational perspective, constraints on the Rastall parameter $\epsilon$ have been obtained across a wide range of physical systems. Analyses of compact stellar structures typically constrain $\epsilon$ at the level $\mathcal{O}(10^{-2})$ \cite{Oliveira2015, ElHanafy2022, ElHanafy2023}, while galaxy--galaxy strong gravitational lensing studies have reported larger values of the order $\mathcal{O}(10^{-1})$ \cite{Li2019}. At the cosmological level, recent analyses combining background expansion and linear perturbations in the flat case indicate that $\epsilon$ is strongly constrained to be very close to zero, with deviations at most of order $\mathcal{O}(10^{-4})$ \cite{Akarsu2020}.

Taken together, these results indicate that the allowed magnitude of the Rastall parameter is strongly dependent on the observational scale and the underlying datasets. This scale dependence motivates further tests of Rastall cosmology using different observational probes and dataset combinations.

Among the outstanding issues in modern cosmology, the tension in measurements of the Hubble constant $H_0$ remains one of the most persistent challenges. In recent years, this discrepancy has become increasingly prominent due to the divergence between two independent measurement approaches. The Planck collaboration, based on observations of the cosmic microwave background (CMB), has reported a value of $H_0 = 67.4 \pm 0.5$ km s$^{-1}$ Mpc$^{-1}$ \cite{aghanim2020planck, Planck:2013pxb, Planck:2015fie}. In contrast, the SH0ES (Supernovae $H_0$ for the Equation of State) team, utilizing distance ladder measurements with the Hubble Space Telescope (HST), has obtained a significantly higher value of $H_0 = 73.2 \pm 1.3$ km s$^{-1}$ Mpc$^{-1}$ \cite{riess2021large}. This discrepancy, often referred to as the "Hubble tension", has profound implications for our understanding of the expansion history of the Universe and may indicate new physics beyond the standard cosmological model \cite{Riess:2019cxk, riess2021large, CAPOZZIELLO2023101201}.

The HST method involves measurements of the local distance ladder by combining photometry of Cepheid variables (via their period–luminosity relation) with other local distance anchors, such as Milky Way parallaxes and calibration distances to Cepheids in nearby galaxies hosting Type Ia supernovae. Specifically, for estimating the Hubble constant, observations of Cepheids in the Large Magellanic Cloud are employed by the HST group \cite{riess2021large}. Currently, the discrepancy between $H_0$ estimations by the Hubble Space Telescope and those inferred from Planck exceeds $4\sigma$. This tension has persisted over the years, as summarized in Table~\ref{table1}. The problem can be interpreted as a discrepancy between early-Universe and late-Universe cosmological observations. The HST group relies on late-time data, while the Planck collaboration combines observations over a wide redshift range ($0 < z < 1100$), adopting the $\Lambda$CDM model as the fiducial cosmological framework.

Different methods have been proposed to address the $H_0$ tension. Several studies have examined $H_0$ estimations using approaches independent of both the Cepheid distance scale and CMB anisotropies \cite{verde2019tensions, di2021snowmass2021}. Among these, the following methods have provided noteworthy results:
\begin{itemize}
	\item Strong gravitational lensing time-delay measurements (e.g., the H0LiCOW project) \cite{wong2020h0licow, denzel2021hubble}.
	\item CMB lensing analyses \cite{wong2020h0licow}.
	\item The tip of the red giant branch (TRGB) method used by the Carnegie–Chicago Hubble Program (CCHP) \cite{freedman2019carnegie, freedman2020calibration}.
	\item Megamaser-hosting galaxies \cite{baxter2021determining}.
	\item Oxygen-rich variable stars (Miras) \cite{Pesce_2020}.
\end{itemize}

Some studies have suggested that either Planck or HST measurements may be affected by systematic uncertainties; however, no consensus explanation has yet resolved the discrepancy.
\begin{table}[h]
	\centering
	\caption{Estimations of $H_0$ (km\,s$^{-1}$\,Mpc$^{-1}$) from different observational methods.}
	\label{table1}
	\setlength{\tabcolsep}{7pt}
	\renewcommand{\arraystretch}{1.4}
	\begin{adjustbox}{max width=\textwidth}
		\begin{tabular}{lcccc}
			\toprule[0.4mm]
			$H_0$ & Project & Method & Year & Ref. \\
			\hline
			$67.4 \pm 0.5$   & Planck  & CMB Power Spectra-Lensing  & 2018  & \cite{aghanim2020planck} \\
			$73.2 \pm 1.3$   & SH0ES (HST)  & Cepheid distance ladder & 2019  & \cite{riess2021large} \\
			$72.7 \pm 4.6$    & HST & 6 Miras in SN Ia host galaxy & 2019  & \cite{huang2020hubble} \\
			$73.3^{+1.7}_{-1.8}$   & HoLiCOW & 6 strong lenses and $\Lambda$CDM  & 2019  & \cite{wong2020h0licow} \\
			$69.6^{+0.8}_{-1.7}$    & CCHP  & TRGB & 2020  & \cite{freedman2020calibration} \\
			$73.9 \pm 3.0$    & 6 maser galaxies  & Megamaser & 2020  & \cite{Pesce_2020} \\
			$68.0 \pm 0.36$    & ACT DR6  & CMB Power Spectra (TT, TE, EE) & 2025  & \cite{ACT_DR6_LCDM} \\
			$71.0 \pm 1.1$    & ACT DR6 + DESI DR2  & CMB + BAO + EDE model fit & 2025  & \cite{ACT_DESI_EDE2025} \\
			\bottomrule[0.4mm]
		\end{tabular}
	\end{adjustbox}
\end{table}

As shown in Table~\ref{table1}, the reported values of the Hubble constant vary between the Planck 2018 result and the SH0ES measurement. Recent late-Universe measurements tend to favor values closer to the SH0ES determination, while early-Universe probes based on the CMB remain consistent with the lower Planck value.

In light of this persistent $H_0$ tension, various extensions to the standard cosmological model have been proposed. These include modified gravity theories \cite{ballesteros2020h0, zumalacarregui2020gravity, Odintsov:2020qzd}, alternative dark energy scenarios \cite{huang2016dark}, frameworks involving additional relativistic species \cite{d2018hot}, and models with interactions between dark energy and dark matter \cite{di2017can}.

In this paper, we conduct a detailed investigation of the Hubble tension within the context of Rastall gravity. This framework introduces a generalized conservation law for the energy-momentum tensor, allowing for an effective interaction between matter fields and spacetime geometry. By applying the Rastall hypothesis in a cosmological setting, we analyze how such modifications affect the expansion history of the Universe and whether the resulting dynamics can alleviate the $H_0$ tension. This approach enables us to confront modified gravity parameters directly with current cosmological observations.

This paper is organized as follows. In Section~\ref{Cosmological-Models}, we describe the cosmological models and parameters considered in this study. The observational datasets are presented in Section~\ref{Observational-data-analysis}. Section~\ref{Constraints-on-Models} outlines the statistical methodology used for model comparison and parameter estimation. The results, including constraints on cosmological parameters and implications for the $H_0$ tension, are discussed in Section~\ref{Results}. Finally, Section~\ref{Conclusions} summarizes the main findings and outlines possible directions for future research. Appendix~\ref{appendix:tension-estimator} provides the statistical estimator used to quantify the tension between datasets.

\section{Cosmological Models}\label{Cosmological-Models}

To determine the Hubble parameter $H(z)$ in different cosmological models, we begin with Einstein’s field equations in general relativity
\begin{equation}\label{1b}
	R_{\mu \nu}-\dfrac{1}{2}Rg_{\mu \nu}=8\pi G  T_{\mu \nu},
\end{equation}
where $g_{\mu \nu}$ is the metric tensor, $G$ is Newton's gravitational constant, and throughout this section we adopt natural units with $8\pi G = 1$. The tensors $R_{\mu \nu}$ and $R$ denote the Ricci tensor and Ricci scalar, respectively. The energy-momentum tensor $T_{\mu \nu}$ describes the total energy content of the universe, including matter, radiation, and dark energy, and is defined as
$$T_{\mu\nu} = \sum_i T^{i}_{\mu\nu},$$
where the index $i$ labels the individual components: baryonic matter and cold dark matter (collectively denoted as $i = m$), radiation ($i = r$), and dark energy ($i = \Lambda$). Each component is modeled as a perfect fluid, characterized by the energy-momentum tensor
\begin{equation}\label{2b}
	T_i^{\mu\nu} = (\rho_i + p_i)u^{\mu}u^{\nu} + p_i g^{\mu\nu},
\end{equation}
where $\rho_i$ and $p_i$ are the energy density and pressure of the $i$-th fluid, and $u^\mu$ is the four-velocity. The conservation of energy-momentum, $T^{\mu}_{\nu;\mu} = 0$, is compatible with the Bianchi identities, $G^{\mu}_{\nu;\mu} = 0$ \cite{Petrov:2020wgy}.
Assuming spatial homogeneity and isotropy, the Einstein equations in a non-flat geometry reduce to the Friedmann equations
\begin{equation}\label{3b}
	H^2+\frac{k}{a^2}= \dfrac{1}{3}\sum_i\rho_i,
\end{equation}
\begin{equation}\label{4b}
	\dfrac{\ddot{a}}{a} = -\dfrac{1}{6}\sum_i(\rho_i + 3p_i) .
\end{equation}
The Hubble parameter is defined as $H \equiv \dot{a}/a$, where $a(t)$ is the cosmological scale factor and the overdot denotes differentiation with respect to cosmic time. The curvature contribution can be formally rewritten as an effective energy density, $\rho_k \equiv -3k/a^2$, where $k = 0$, $+1$, and $-1$ correspond to flat, closed, and open universes, respectively.
The evolution of each fluid component is determined by its equation of state, given by $p_i = \omega_i \rho_i$, where $\omega_i$ is the equation of state (EoS) parameter. For pressureless matter and cold dark matter, $\omega_m = 0$; for radiation, $\omega_r = 1/3$; for a cosmological constant, $\omega_\Lambda = -1$; and for a generic dark energy component $X$, one has $\omega_X < -1/3$.
In addition, the energy conservation equation is given by
\begin{equation}\label{6b}
	\dot{\rho_i} = -3H(p_i + \rho_i) = -3H(1 + \omega_i)\rho_i.
\end{equation}
Assuming a constant EoS parameter $\omega_i$, the solution to Eq.~(\ref{6b}) is
\begin{equation}\label{7b}
	\rho_i(a) = \rho_{i0} \left(\frac{a_0}{a}\right)^{3(1 + \omega_i)},
\end{equation}
where $\rho_{i0}$ and $a_0$ denote the present values of the energy density of the $i$-th component and the scale factor, respectively. In what follows, we adopt the normalization $a_0 = 1$.
The dimensionless energy density parameters at the present time are defined as
\begin{equation}\label{9b}
	\Omega_{m_0}\equiv\dfrac {\rho_{m_0}}{3H_0^2},\hspace{0.1cm}  
	\Omega_{\Lambda_0} \equiv \dfrac{\rho_{\Lambda_0}}{3H_0^2}, \hspace{0.1cm}
	\Omega_{r_0} \equiv \frac{\rho_{r_0}}{3H_0^2},\hspace{0.1cm}
	\Omega_{k_0} \equiv -\frac{k}{a_0^2H_0^2}.
\end{equation}
Using these definitions, the Friedmann equation (\ref{3b}) for the $\Lambda$CDM model can be written as
\begin{align}\label{10b}
	E(z;\boldsymbol{p}) \equiv \frac{H(z;\boldsymbol{p})}{H_0} = \Big[ \Omega_{m_0} (1+z)^3 + \Omega_{r_0} (1+z)^4 
	+ \Omega_{k_0}(1+z)^2 + \Omega_{\Lambda_{0}} \Big]^{1/2}
\end{align}
where $z= (a_0/a) - 1$ is the redshift, $H_0$ is the present Hubble parameter, and $\boldsymbol{p}$ denotes the set of model parameters. In the flat case, $\Omega_{k_0}=0$, while for open (closed) geometries $\Omega_{k_0}>0$ ($\Omega_{k_0}<0$). The density parameters satisfy the closure condition $\Omega_{\Lambda_0} + \Omega_{m_0} + \Omega_{r_0} + \Omega_{k_0}=1$.
\subsection{Rastall Model}\label{subsec:RastalModel}

A notable approach to modifying General Relativity consists in relaxing the standard conservation law of the energy-momentum tensor. In this framework, the usual condition $T^{\mu}_{\nu;\mu}=0$ is modified, leading to deviations from standard geodesic motion and the appearance of an additional force \cite{Lobo:2022aop}. Rastall originally proposed such a modification by allowing a non-vanishing covariant divergence of the matter energy--momentum tensor \cite{Rastall1972, Rastall1976}.

It is worth noting that the theoretical status of Rastall gravity has been the subject of ongoing discussion. In particular, it has been argued that the theory can be formally recast as standard General Relativity with a redefined non-conserved effective energy--momentum tensor \cite{Visser2017}. On the other hand, other studies have emphasized that, depending on the physical interpretation and the matter sector, Rastall gravity may still lead to phenomenological deviations from GR in specific astrophysical and cosmological contexts \cite{Darabi2018}. 

In the present work, we adopt this phenomenological viewpoint and treat Rastall gravity as an effective framework for probing deviations from the standard conservation law at cosmological scales. Within this approach, the parameter $\epsilon$ can be directly constrained by cosmological observations, as discussed in the following sections.

The aim here is to place constraints on cosmological parameters in both the standard $\Lambda$CDM model and its generalized counterpart, referred to as the R--$\Lambda$CDM model, in order to investigate their effectiveness in alleviating the $H_0$ tension. In Rastall gravity, the modified Einstein field equations are written as
\begin{equation}\label{12b}
	G_{\mu\nu}
	=
	\kappa \left( T_{\mu\nu} + \hat{T}_{\mu\nu} \right),
\end{equation}
where
\begin{equation*}
	\hat{T}_{\mu\nu} = -\epsilon R g_{\mu\nu}.
\end{equation*}
The effective gravitational coupling $\kappa$ satisfies
\begin{equation}\label{kappa}
	\kappa
	=
	8\pi G
	\left(
	\frac{1-4\epsilon \kappa}
	{1-6\epsilon \kappa}
	\right).
\end{equation}
Adopting $8\pi G = 1$, Eq.~(\ref{kappa}) reduces to
\begin{equation}\label{13b}
	6\epsilon \kappa^2
	-
	(1+4\epsilon)\kappa
	+
	1
	=
	0.
\end{equation}
The requirement that $\kappa$ be real implies
\begin{equation}\label{kappa_condition}
	\epsilon
	\le
	\frac{2-\sqrt{3}}{4}
	\simeq 0.067
	\quad
	\text{or}
	\quad
	\epsilon
	\ge
	\frac{2+\sqrt{3}}{4}.
\end{equation}
We discard the second branch as it corresponds to large deviations from general relativity. In what follows, we focus on the phenomenologically relevant regime
\begin{equation}\label{small_eps}
	|\epsilon \kappa| \ll 1,
\end{equation}
which motivates the restriction
\begin{equation}\label{14b}
	\epsilon \in (-0.10,\,0.07).
\end{equation}
Under the perturbative branch continuously connected to general relativity, expanding Eq.~(\ref{13b}) around $\kappa=1$ yields
\begin{equation}\label{kappa_expansion}
	\kappa \simeq 1 + 2\epsilon.
\end{equation}
The field equations then take the approximate form
\begin{equation}\label{15b}
	G_{\mu\nu}
	=
	T_{\mu\nu}
	+
	\epsilon
	\left[
	2 T_{\mu\nu}
	-
	(1+2\epsilon) R g_{\mu\nu}
	\right].
\end{equation}
Taking the trace of the Rastall field equations yields the exact relation
\begin{equation}\label{trace_exact}
	R
	=
	-\frac{\kappa}{1-4\epsilon\kappa}\,T.
\end{equation}
Using the perturbative solution given in Eq.~(\ref{kappa_expansion}), the Ricci scalar becomes
\begin{equation}\label{20b}
	R
	=
	-
	(1+2\epsilon)
	\left(
	1 + 4\epsilon + 24\epsilon^2 + \cdots
	\right)
	T ,
\end{equation}
where we consistently retain terms up to $\mathcal{O}(\epsilon^2)$ throughout the cosmological analysis.
Substituting back, we obtain
\begin{equation}\label{17b}
	G^{\mu}_{\nu}
	=
	T^{\mu}_{\nu}
	+
	\epsilon
	\left[
	2 T^{\mu}_{\nu}
	+
	(1+8\epsilon) g^{\mu}_{\nu} T
	\right].
\end{equation}
For a perfect fluid,
\begin{equation*}
	{T_{(i)}}^{\mu}_{\nu}
	=
	\mathrm{diag}(-\rho_i, p_i, p_i, p_i),
\end{equation*}
with equation-of-state parameter $\omega_i = p_i/\rho_i$. In a non-flat FLRW background, the modified Friedmann equation becomes
\begin{equation}\label{18b}
	H^2 + \frac{k}{a^2}
	=
	\frac{1}{3}
	\sum_i
	\rho_i
	\left[
	1
	+ 3\epsilon(1-\omega_i)
	+ 8\epsilon^2(1-3\omega_i)
	\right].
\end{equation}
The corresponding modified acceleration equation can be written as
\begin{equation}\label{acceleration_rastall_exact}
	\frac{\ddot a}{a}
	=
	\frac{
		\left[1-6\epsilon(1+2\epsilon)\right]
		\frac{1}{3}
		\sum_i
		\rho_i
		\left[
		1+3\epsilon(1-\omega_i)
		+8\epsilon^2(1-3\omega_i)
		\right]
		+
		(1+2\epsilon)
		\sum_i \omega_i\rho_i
	}{
		6\epsilon(1+2\epsilon)-2
	}.
\end{equation}
Expanding Eq.~(\ref{acceleration_rastall_exact}) up to $\mathcal{O}(\epsilon^2)$ yields
\begin{equation}\label{acceleration_rastall_expanded}
	\frac{\ddot a}{a}
	=
	-\frac{1}{6}
	\sum_i
	\rho_i
	\left[
	1+3\omega_i
	+12\epsilon\omega_i
	+16\epsilon^2(3\omega_i-1)
	\right].
\end{equation}
In the limit $\epsilon\to0$, Eq.~(\ref{acceleration_rastall_expanded}) reduces to the standard GR acceleration equation,
\begin{equation}
	\frac{\ddot a}{a}
	=
	-\frac{1}{6}
	\sum_i
	(\rho_i+3p_i).
\end{equation}
The modified continuity equation reads
\begin{equation}\label{continuity}
	\dot{\rho}_i
	=
	-3\frac{\dot{a}}{a}
	\left[
	\frac{1+\omega_i}
	{1+\epsilon(1+6\epsilon)(3\omega_i-1)}
	\right]
	\rho_i,
\end{equation}
whose solution is
\begin{equation}\label{19b}
	\rho_i
	=
	\rho_{i0}
	\left(
	\frac{a_0}{a}
	\right)^{
		\frac{3(1+\omega_i)}
		{1+\epsilon(1+6\epsilon)(3\omega_i-1)}
	}.
\end{equation}
Equation~(\ref{19b}) shows that the Rastall parameter $\epsilon$ modifies the effective dilution rate of non-relativistic matter relative to the standard $\Lambda$CDM scenario. In particular, for $\epsilon>0$ the matter density evolves faster than the standard $(1+z)^3$ scaling, leading to a mild suppression of late-time matter clustering, whereas $\epsilon<0$ produces a slower matter dilution and therefore tends to enhance clustering relative to GR. In contrast, radiation and vacuum energy preserve their standard evolution laws, namely $\rho_r\propto(1+z)^4$ and $\rho_\Lambda=\mathrm{const}$. Consequently, the main cosmological effect of Rastall gravity at the background level arises through the modified matter evolution.
We adopt $a_0=1$ and consistently retain terms up to $\mathcal{O}(\epsilon^2)$. For convenience, we define the rescaled density parameters
\begin{equation}\label{OmegaiNew}
	\tilde{\Omega}_i
	=
	\Omega_i
	\left[
	1
	+ 3\epsilon(1-\omega_i)
	+ 8\epsilon^2(1-3\omega_i)
	\right],
\end{equation}
which govern the background expansion. The Hubble expansion rate is therefore
\begin{equation}\label{21b}
	\frac{H(z;\boldsymbol{p})}{H_0}
	=
	\Big[
	\tilde{\Omega}_{m_0}
	(1+z)^{\frac{3}{1-\epsilon(1+6\epsilon)}}
	+ \tilde{\Omega}_{r_0} (1+z)^4
	+ \Omega_{k_0} (1+z)^2
	+ \tilde{\Omega}_{\Lambda_0}
	\Big]^{1/2}.
\end{equation}
with
\begin{align}\label{22b}
	\tilde{\Omega}_{m_0}
	&=
	\Omega_{m_0}(1+3\epsilon+8\epsilon^2), \\
	\tilde{\Omega}_{r_0}
	&=
	\Omega_{r_0}(1+2\epsilon), \\
	\tilde{\Omega}_{\Lambda_0}
	&=
	\Omega_{\Lambda_0}(1+6\epsilon+32\epsilon^2),
\end{align}
and the closure condition
\begin{equation}\label{closure}
	\tilde{\Omega}_{\Lambda_0}
	+
	\tilde{\Omega}_{m_0}
	+
	\tilde{\Omega}_{r_0}
	+
	\Omega_{k_0}
	=
	1.
\end{equation}
We refer to these quantities as effective density parameters, $\Omega_i^{\rm eff} \equiv \tilde{\Omega}_i$, which reduce to the standard $\Lambda$CDM values in the limit $\epsilon \to 0$.

\section{Observational data analysis}\label{Observational-data-analysis}

In the following, we derive constraints on the parameters of both the standard and Rastall versions of the $\Lambda$CDM model. To this end, we apply the Markov Chain Monte Carlo (MCMC) method to explore the parameter space and maximize the likelihood function $L(\boldsymbol{p})$, or equivalently minimize $\chi^2(\boldsymbol{p})$ under the assumption of Gaussian likelihoods \cite{farooq2017hubble}, with respect to the parameter vector $\boldsymbol{p}$, in order to determine the best-fit values $\boldsymbol{p}_0$. Confidence regions at the $1\sigma$, $2\sigma$, and $3\sigma$ levels are presented in terms of two-dimensional parameter contours.

For the calculation of $\chi^2(\boldsymbol{p})$, we use the following observational datasets:
\begin{itemize}
	\item Big Bang Nucleosynthesis (BBN) constraints based on the updated PRIMAT framework \cite{Pitrou2018,PRIMAT},
	\item $H(z)$ measurements, incorporating 40 data points as listed in Table~\ref{hubble},
	\item Type Ia Supernovae data from the Pantheon+ compilation, including the SH0ES calibration and the full STAT+SYS covariance matrix \cite{Brout_2022},
	\item CMB distance priors from Planck 2018, using the appropriate flat or curved priors and the associated covariance matrix \cite{aghanim2020planck,Chen:2018dbv},
	\item BAO measurements from the DESI DR2 Gaussian BAO dataset (GCcomb), employing the full mean vector and covariance matrix \cite{DESI2025_DR2_II},
	\item Growth-rate measurements $f(z)\sigma_8(z)$, consisting of 26 data points as listed in Table~\ref{fs8}.
\end{itemize}
Assuming statistical independence among the different datasets, the total likelihood for a given data combination is given by the product of the corresponding likelihood terms,
\begin{equation}
	L_{\rm tot}(\boldsymbol{p}) \propto 
	L_{\rm BBN} \times L_{\rm H} \times L_{\rm SN} \times L_{\rm CMB} \times L_{\rm BAO} \times L_{f\sigma_8},
\end{equation}
and therefore the combined chi-square can be written as
\begin{align}\label{26b} 	
	\chi_{\rm tot}^2(\boldsymbol{p}) =
	\chi_{\rm BBN}^2(\boldsymbol{p})
	+ \chi_{\rm H}^2(\boldsymbol{p})
	+ \chi_{\rm SN}^2(\boldsymbol{p})
	+ \chi_{\rm CMB}^2(\boldsymbol{p})
	+ \chi_{\rm BAO}^2(\boldsymbol{p})
	+ \chi_{f\sigma_8}^2(\boldsymbol{p}),
\end{align}
where each contribution is described in detail in the following subsections.

\subsection{Big Bang Nucleosynthesis (BBN data)}

Big Bang Nucleosynthesis (BBN) provides an independent constraint on the physical baryon density through the primordial light-element abundances. In this work, we employ the publicly available PRIMAT framework \cite{Pitrou2018}, using the most recent grid release (September 2024 update)\footnote{PRIMAT public grid: \url{https://www2.iap.fr/users/pitrou/primat.htm}}.
The PRIMAT calculation includes QED corrections and incomplete neutrino decoupling effects \cite{Froustey2020}, such that $\Delta N = 0$ corresponds to an effective number of relativistic species $N_{\rm eff} \simeq 3.044$. We fix $N_{\rm eff}=3.046$ throughout the analysis, which is fully consistent with the PRIMAT prediction for $\Delta N=0$ and with the standard value adopted in Planck 2018 CMB analyses.

The theoretical prediction for the primordial helium-4 mass fraction, $Y_p^{\rm th}$, is obtained by interpolating the PRIMAT grid as a function of the physical baryon density $\omega_b \equiv \Omega_{b0}h^2$ and $\Delta N$, and evaluated at $\Delta N=0$. We compare this prediction with the observational determination $Y_p^{\rm obs}= 0.2448 \pm 0.0033$ \cite{Aver2022}, assuming a Gaussian uncertainty, and define the BBN contribution to the total chi-square as

\begin{equation}
	\chi^2_{\rm BBN} =
	\left[
	\frac{Y_p^{\rm th}(\Omega_{b0}h^2,\Delta N=0)-Y_p^{\rm obs}}
	{\sigma_{Y_p}}
	\right]^2 ,
\end{equation}
where $\sigma_{Y_p}= 0.0033$.
This BBN likelihood provides an observationally driven and independent constraint on the baryon density, complementary to late-time cosmological probes.

\subsection{Hubble parameter measurements ($H(z)$ data)}

We use 40 independent measurements of the Hubble parameter obtained from the cosmic chronometer (CC) method, as summarized in Table~\ref{hubble}, spanning the redshift range $0.07 \leq z \leq 2.3$. These measurements are based on the differential age evolution of passively evolving galaxies and provide direct determinations of the expansion rate. No direct $H_0$ prior at $z=0$ is included in this compilation.

We emphasize that none of the $H(z)$ measurements used in this work are derived from baryon acoustic oscillation (BAO) techniques. Therefore, the $H(z)$ dataset is statistically independent of the DESI BAO measurements included in our analysis, and no overlap or double counting is present.

Assuming no correlations among the measurements, the contribution of the $H(z)$ data to the total chi-square is defined as
\begin{equation} \label{28b}
	\chi_{\rm H}^2(\boldsymbol{p}) =
	\sum_{i=1}^{40}
	\frac{\left[ H_{\rm th}(z_i; \boldsymbol{p}) - H_{\rm obs}(z_i) \right]^2}{\sigma_i^2},
\end{equation}
where $H_{\rm th}(z_i; \boldsymbol{p}) = H_0\,E(z_i;\boldsymbol{p})$ denotes the theoretical prediction of the Hubble parameter at redshift $z_i$.

This dataset provides a direct probe of the late-time expansion history and plays a central role in constraining the Hubble constant within both the $\Lambda$CDM and R-$\Lambda$CDM frameworks.

\begin{table}[h]
	\caption{Compilation of 40 independent measurements of the Hubble parameter $H(z)$ (in ${\rm km\,s^{-1}\,Mpc^{-1}}$), covering the redshift range $0.07 \leq z \leq 2.3$.}
	\centering
	\label{hubble}
	\setlength{\tabcolsep}{7pt}
	\renewcommand{\arraystretch}{1.2}
	\begin{adjustbox}{max width=\textwidth}
		\begin{tabular}{lccc|lccc}
			\toprule[0.4mm]
			$z$ & $H(z)$ & $\sigma_H$ & Ref.
			& $z$ & $H(z)$ & $\sigma_H$ & Ref. \\
			\hline
			0.070 & 69.00 & 19.60 & \cite{Zhang2014}
			& 0.480 & 97.00 & 62.00 & \cite{Stern2010} \\
			
			0.090 & 69.00 & 12.00 & \cite{Simon2005}
			& 0.500 & 72.10 & 34.60 & \cite{Loubser2025} \\
			
			0.100 & 69.00 & 12.00 & \cite{Stern2010}
			& 0.570 & 89.20 & 3.60 & \cite{Moresco2012} \\
			
			0.120 & 68.60 & 26.20 & \cite{Zhang2014}
			& 0.593 & 104.00 & 13.00 & \cite{Moresco2012} \\
			
			0.170 & 83.00 & 8.00 & \cite{Stern2010}
			& 0.680 & 92.00 & 8.00 & \cite{Moresco2012} \\
			
			0.180 & 75.00 & 4.00 & \cite{Moresco2012}
			& 0.730 & 97.30 & 7.00 & \cite{Moresco2012} \\
			
			0.200 & 72.90 & 29.60 & \cite{Zhang2014}
			& 0.750 & 105.00 & 10.80 & \cite{Jimenez2023} \\
			
			0.240 & 79.69 & 2.65 & \cite{Simon2005}
			& 0.753 & 98.80 & 33.60 & \cite{Borghi_2022} \\
			
			0.270 & 77.00 & 14.00 & \cite{Moresco2012}
			& 0.800 & 113.10 & 32.50 & \cite{Jiao2022} \\
			
			0.280 & 88.80 & 36.60 & \cite{Zhang2014}
			& 0.875 & 125.00 & 17.00 & \cite{Moresco2012} \\
			
			0.350 & 82.10 & 4.80 & \cite{Moresco2016}
			& 0.900 & 117.00 & 23.00 & \cite{Stern2010} \\
			
			0.352 & 83.00 & 14.00 & \cite{Moresco2016}
			& 1.037 & 154.00 & 20.00 & \cite{Moresco2012} \\
			
			0.380 & 83.00 & 13.50 & \cite{Moresco2016}
			& 1.260 & 135.00 & 65.00 & \cite{Tomasetti2023} \\
			
			0.400 & 77.00 & 10.20 & \cite{Moresco2016}
			& 1.300 & 168.00 & 17.00 & \cite{Moresco2012} \\
			
			0.425 & 87.10 & 11.20 & \cite{Moresco2016}
			& 1.363 & 160.00 & 33.60 & \cite{Moresco2015} \\
			
			0.429 & 91.80 & 5.30 & \cite{Moresco2016}
			& 1.430 & 177.00 & 18.00 & \cite{Moresco2015} \\
			
			0.440 & 82.60 & 7.80 & \cite{Stern2010}
			& 1.530 & 140.00 & 14.00 & \cite{Stern2010} \\
			
			0.450 & 92.80 & 12.90 & \cite{Moresco2016}
			& 1.750 & 202.00 & 40.00 & \cite{Stern2010} \\
			
			0.470 & 89.00 & 34.00 & \cite{Ratsimbazafy2017}
			& 1.965 & 186.50 & 50.40 & \cite{Moresco2015} \\
			
			0.478 & 80.90 & 9.00 & \cite{Moresco2016}
			& 2.300 & 224.00 & 8.00 & \cite{Moresco2015} \\
			\bottomrule[0.4mm]
		\end{tabular}
	\end{adjustbox}
\end{table}

\subsection{Type Ia supernovae: Pantheon+SH0ES sample (SN data)}

Type Ia supernovae (SNe Ia) provide precise measurements of the late-time expansion history through the luminosity distance–redshift relation. In this work, we employ the Pantheon+SH0ES compilation \cite{Brout_2022}, which consists of 1701 light curves of 1555 spectroscopically confirmed SNe Ia in the redshift range $0.001 < z < 2.3$. This sample includes Cepheid-calibrated distances from the SH0ES program and provides observed distance moduli together with the full statistical and systematic covariance matrix.
The theoretical distance modulus is defined as
\begin{equation}
	\mu_{\rm th}(z) = 5 \log_{10} \left( \frac{d_L(z)}{\rm Mpc} \right) + 25 ,
\end{equation}
where the luminosity distance is
\begin{equation}
	d_L(z) = (1+z)\, S_k(\chi),
\end{equation}
with
\begin{equation}
	\chi = \frac{c}{H_0} \int_0^z \frac{dz'}{E(z')},
	\qquad
	E(z) \equiv \frac{H(z)}{H_0},
\end{equation}
where $c$ denotes the speed of light. The transverse comoving distance function is given by
\begin{equation}
	S_k(\chi) =
	\begin{cases}
		\frac{1}{\sqrt{\Omega_{k0}}}
		\sinh\!\left(\sqrt{\Omega_{k0}}\,\chi\right),
		& \Omega_{k0} > 0, \\[6pt]
		\chi,
		& \Omega_{k0} = 0, \\[6pt]
		\frac{1}{\sqrt{-\Omega_{k0}}}
		\sin\!\left(\sqrt{-\Omega_{k0}}\,\chi\right),
		& \Omega_{k0} < 0 .
	\end{cases}
\end{equation}
Distances are evaluated at the CMB-frame redshifts provided in the Pantheon + SH0ES catalogue. All observational uncertainties, including peculiar velocity corrections and calibration systematics, are incorporated through the full covariance matrix ${\sf C}$ (STAT+SYS). The absolute magnitude parameter $M_B$ is treated as a nuisance parameter and analytically marginalized over, such that the supernova data do not impose an explicit prior on $H_0$.
The supernova chi-square is defined as
\begin{equation}
	\chi^2_{\rm SN} =
	(\boldsymbol{\mu}_{\rm th} - \boldsymbol{\mu}_{\rm obs})^{T}
	{\sf C}^{-1}
	(\boldsymbol{\mu}_{\rm th} - \boldsymbol{\mu}_{\rm obs}) .
\end{equation}
Following the standard procedure \cite{Conley_2011}, we analytically marginalize over the nuisance parameter associated with the absolute magnitude $M_B$. Defining
\begin{equation}
	\boldsymbol{\Delta}_0 =
	\boldsymbol{\mu}_{\rm th}\big|_{M_B=0} - \boldsymbol{\mu}_{\rm obs},
\end{equation}
the marginalized chi-square can be written as
\begin{equation}
	\chi^2_{\rm SN} = a - \frac{b^2}{c},
\end{equation}
where
\begin{equation}
	a = \boldsymbol{\Delta}_0^{T} {\sf C}^{-1} \boldsymbol{\Delta}_0,
	\quad
	b = \mathbf{1}^{T} {\sf C}^{-1} \boldsymbol{\Delta}_0,
	\quad
	c = \mathbf{1}^{T} {\sf C}^{-1} \mathbf{1},
\end{equation}
and $\mathbf{1}$ denotes a vector of ones with dimension equal to the number of supernovae.
For numerical stability and efficiency, the inverse-covariance operations are implemented using a Cholesky decomposition of the covariance matrix.

\subsection{Cosmic Microwave Background (CMB data)}

We employ the CMB distance priors derived from Planck 2018 data, as provided in \cite{Chen:2018dbv}. The compressed CMB information is encoded in the shift parameter $R$, the acoustic scale $l_A$, and the physical baryon density $\Omega_{b0}h^2$.
The shift parameter is defined as
\begin{equation}
	R = \sqrt{\Omega_{m0}} \frac{H_0}{c} D_M(z_*),
\end{equation}
where $D_M(z)$ denotes the transverse comoving distance evaluated consistently with the assumed spatial curvature, and $c$ is the speed of light. The acoustic scale is given by
\begin{equation}
	l_A = \pi \frac{D_M(z_*)}{r_s(z_*)},
\end{equation}
where $r_s(z_*)$ is the comoving sound horizon at the redshift of recombination $z_*$. The sound horizon is computed as
\begin{equation}
	r_s(a) = \int_{0}^{a} \frac{c_s(a')}{a'^2 H(a')} \, da',
\end{equation}
with the baryon sound speed
\begin{equation}
	c_s(a) = \frac{1}{\sqrt{3\left(1 + a \frac{3\Omega_{b0}}{4\Omega_{r0}}\right)}}.
\end{equation}
Throughout this analysis, we assume standard radiation content with $N_{\rm eff}=3.046$. The recombination redshift $z_*$ is evaluated using the fitting formula of \cite{hu1996small}. The CMB chi-square is defined as
\begin{equation}
	\chi_{\rm CMB}^2 = X_{\rm CMB}^T C_{\rm CMB}^{-1} X_{\rm CMB},
\end{equation}
where $X_{\rm CMB}$ depends on the assumed spatial geometry. For spatially flat models, we adopt
\begin{equation}
	X_{\rm CMB}^{\rm flat} =
	\begin{pmatrix}
		R - 1.7502 \\
		l_A - 301.471 \\
		\Omega_{b0}h^2 - 0.02236
	\end{pmatrix},
\end{equation}
with inverse covariance matrix
\begin{equation}
	C_{\rm CMB,flat}^{-1} =
	\begin{pmatrix}
		9.43923971\times10^{4} & -1.36049130\times10^{3} & 1.6645172916\times10^{6} \\
		-1.36049130\times10^{3} & 1.61434900\times10^{2} & 3.67161800\times10^{3} \\
		1.6645172916\times10^{6} & 3.67161800\times10^{3} & 7.97191825162\times10^{7}
	\end{pmatrix}.
\end{equation}

For non-flat (open or closed) models, we use
\begin{equation}
	X_{\rm CMB}^{\rm curved} =
	\begin{pmatrix}
		R - 1.7429 \\
		l_A - 301.409 \\
		\Omega_{b0}h^2 - 0.02260
	\end{pmatrix},
\end{equation}
with inverse covariance matrix
\begin{equation}
	C_{\rm CMB,curved}^{-1} =
	\begin{pmatrix}
		1.02243282\times10^{5} & -1.56541591\times10^{3} & 1.94853151\times10^{6} \\
		-1.56541591\times10^{3} & 1.70590196\times10^{2} & 3.13083183\times10^{3} \\
		1.94853151\times10^{6} & 3.13083183\times10^{3} & 7.91479197\times10^{7}
	\end{pmatrix}.
\end{equation}
Since the allowed values of the Rastall parameter correspond to small deviations from general relativity and affect only the background expansion at the perturbative level considered here, we consistently adopt the Planck 2018 distance priors derived within the standard cosmological framework.

\subsection{Baryon Acoustic Oscillations (BAO data)}

We use the Baryon Acoustic Oscillation (BAO) measurements from the Dark Energy Spectroscopic Instrument (DESI) Data Release 2 (DR2), based on three years of observations \cite{DESI2025_DR2_II,DESI2025_DR2_I,Lodha2025}. These data include post-reconstruction clustering measurements from over 14 million galaxies and quasars, providing 13 distance constraints over the redshift range $0.295 < z < 2.33$. The DESI DR2 sample includes:
(i) isotropic measurements of $D_V(z)/r_s(z_d)$ at low redshift, and
(ii) anisotropic measurements of $D_M(z)/r_s(z_d)$ and $D_H(z)/r_s(z_d)$
at higher redshift,
where $r_s(z_d)$ is the comoving sound horizon at the baryon drag epoch. The transverse comoving distance is defined as
\begin{equation}
	D_M(z) =
	S_k \!\left(
	\frac{c}{H_0}
	\int_0^z \frac{dz'}{E(z')}
	\right),
\end{equation}
where $E(z)=H(z)/H_0$ and $c$ is the speed of light. The curvature-dependent function $S_k(\chi)$ is given by
\begin{equation}
	S_k(\chi) =
	\begin{cases}
		\frac{1}{\sqrt{\Omega_{k0}}}
		\sinh\!\left(\sqrt{\Omega_{k0}}\,\chi\right),
		& \Omega_{k0} > 0, \\[6pt]
		\chi, & \Omega_{k0}=0, \\[6pt]
		\frac{1}{\sqrt{-\Omega_{k0}}}
		\sin\!\left(\sqrt{-\Omega_{k0}}\,\chi\right),
		& \Omega_{k0} < 0 .
	\end{cases}
\end{equation}
The Hubble distance is
\begin{equation}
	D_H(z) = \frac{c}{H(z)} .
\end{equation}
The volume-averaged distance used for isotropic BAO measurements is
\begin{equation}
	D_V(z) = \left[ z\, D_M^2(z)\, D_H(z) \right]^{1/3},
\end{equation}
which is equivalent to the standard definition expressed in terms of the angular diameter distance. The sound horizon at the drag epoch, $r_s(z_d)$, is computed consistently with the background cosmology, assuming $N_{\rm eff}=3.046$, and using the same radiation content as in the CMB analysis. The BAO likelihood is assumed to be Gaussian in the measured distance ratios relative to $r_s(z_d)$. The corresponding chi-square is defined as
\begin{equation}
	\chi^2_{\rm BAO}
	=
	(\mathbf{d}_{\rm obs}-\mathbf{d}_{\rm th})^T
	\mathbf{C}^{-1}
	(\mathbf{d}_{\rm obs}-\mathbf{d}_{\rm th}),
\end{equation}
where $\mathbf{d}_{\rm obs}$ is the vector of the 13 DESI DR2 measurements ($D_V/r_s$, $D_M/r_s$, or $D_H/r_s$ depending on the redshift bin), $\mathbf{d}_{\rm th}$ is the corresponding theoretical prediction, and $\mathbf{C}$ is the full $13\times13$ covariance matrix provided by the DESI collaboration for the combined galaxy+quasar sample (excluding Ly$\alpha$ forest).

Since the CMB distance priors and DESI BAO measurements probe different physical observables and are derived from statistically independent datasets, their likelihoods are treated as independent in the combined analysis.

\subsection{The growth rate of cosmic structures ($f\sigma_8(z)$ data)}

While the geometrical probes discussed in the previous sections primarily constrain the background expansion history, the growth of cosmic structures provides an independent and complementary test of cosmological models at the perturbation level. Measurements of the quantity $f(z)\sigma_8(z)$ directly probe the evolution of matter density fluctuations and are therefore sensitive to both the expansion rate and the underlying theory of gravity.

In this work, we use 26 independent measurements of $f(z)\sigma_8(z)$ compiled in Table~\ref{fs8}. In the absence of a publicly available full covariance matrix for the combined growth sample, the data points are treated as statistically independent and a diagonal covariance matrix is assumed. The corresponding chi-square is therefore written as
\begin{equation}
	\chi_{f\sigma_8}^2
	=
	\sum_i
	\frac{\left[
		f\sigma_{8,\rm th}(z_i)
		-
		f\sigma_{8,\rm obs}(z_i)
		\right]^2}
	{\sigma_i^2}.
\end{equation}
In the linear regime ($\delta \ll 1$), the evolution of matter perturbations in a homogeneous and isotropic background is governed by relativistic perturbation theory. For sub-horizon modes and in the absence of dark-energy clustering, the matter density contrast $\delta_m$ satisfies
\begin{equation}
	\delta_m''(a)
	+
	\left[
	\frac{3}{a}
	+
	\frac{E'(a)}{E(a)}
	\right]
	\delta_m'(a)
	-
	\frac{3}{2}
	\frac{\Omega_{m0}}{a^5 E^2(a)}
	\delta_m(a)
	=
	0,
\end{equation}
where a prime denotes differentiation with respect to the scale factor $a$, and
\begin{equation}
	E(a) = \frac{H(a)}{H_0}.
\end{equation}
The matter density parameter evolves as
\begin{equation}
	\Omega_m(a) = \frac{\Omega_{m0} a^{-3}}{E^2(a)}.
\end{equation}

In the present analysis, we assume homogeneous dark energy with effective sound speed $C_{\rm eff}^2 = 1$ and vanishing dark-energy perturbations, $\delta_\Lambda = 0$, which is well justified on sub-horizon scales where dark-energy fluctuations are strongly suppressed. For the Rastall model, we allow the modification to enter only through the background expansion rate $H(a)$, while the linear perturbation equation retains its standard form. 
In this approach, the effective density parameters $\tilde{\Omega}_i$ are used only in the background expansion history entering $E(a)$, whereas the perturbation source term is written in terms of the standard matter density parameter $\Omega_{m0}$. Therefore, Rastall corrections affect the growth evolution indirectly through the modified Hubble expansion rate.
This approximation is consistent with the perturbative regime $|\epsilon| \ll 1$ considered in this work. The growth rate function and the matter fluctuation amplitude are defined as
\begin{equation}
	f(z)
	=
	\frac{d\ln\delta_m}{d\ln a}
	=
	\frac{\delta_m'(a)}{a\,\delta_m(a)},
\end{equation}
\begin{equation}
	\sigma_8(z)
	=
	\sigma_8(0)
	\frac{\delta_m(z)}{\delta_m(0)},
\end{equation}
where $\sigma_8(0)$ is treated as a free cosmological parameter in the MCMC analysis. The product $f(z)\sigma_8(z)$ constitutes a directly measurable quantity and provides a powerful probe for testing cosmological models at the perturbative level.

\begin{table}[h]
	\caption{The observational function $f(z)\sigma_8(z)$ versus redshift (updated with DESI 2024 Y1 measurements).}
	\centering
	\label{fs8}
	\setlength{\tabcolsep}{7pt}
	\renewcommand{\arraystretch}{1.2}
	\begin{adjustbox}{max width=\textwidth}
		\begin{tabular}{lcccc|lcccc}
			\toprule[0.4mm]
			$z$ & $f\sigma_8$ & $\sigma$ & Year & Ref. 
			& $z$ & $f\sigma_8$ & $\sigma$ & Year & Ref. \\
			\hline
			0.02  & 0.428 & 0.0465 & 2017 & \cite{huterer2017testing} 
			& 0.60  & 0.550 & 0.120 & 2017 & \cite{alam2017clustering} \\
			
			0.025 & 0.390 & 0.110  & 2017 & \cite{achitouv2017_6dfgs_voids} 
			& 0.70  & 0.473 & 0.041 & 2021 & \cite{bautista2021eboss} \\
			
			0.067 & 0.423 & 0.055  & 2012 & \cite{beutler2012_6dfgs} 
			& 0.71  & 0.484 & 0.055 & 2024 & \cite{Adame_2025} \\
			
			0.10  & 0.370 & 0.130  & 2015 & \cite{feix2015growth} 
			& 0.73  & 0.437 & 0.072 & 2012 & \cite{blake2012wigglez} \\
			
			0.15  & 0.530 & 0.160  & 2015 & \cite{howlett2015clustering} 
			& 0.85  & 0.315 & 0.095 & 2021 & \cite{demattia2021eboss} \\
			
			0.17  & 0.510 & 0.060  & 2009 & \cite{song2009reconstructing} 
			& 0.86  & 0.400 & 0.110 & 2017 & \cite{alam2017clustering} \\
			
			0.18  & 0.360 & 0.090  & 2013 & \cite{blake2013galaxy} 
			& 0.92  & 0.422 & 0.046 & 2024 & \cite{Adame_2025} \\
			
			0.295 & 0.378 & 0.094  & 2024 & \cite{Adame_2025} 
			& 1.05  & 0.280 & 0.080 & 2016 & \cite{wilson2016_vipers_v7} \\
			
			0.38  & 0.440 & 0.060  & 2013 & \cite{blake2013galaxy} 
			& 1.32  & 0.375 & 0.039 & 2024 & \cite{Adame_2025} \\
			
			0.38  & 0.497 & 0.045  & 2017 & \cite{alam2017boss} 
			& 1.40  & 0.482 & 0.116 & 2016 & \cite{okumura2016subaru} \\
			
			0.44  & 0.413 & 0.080  & 2012 & \cite{blake2012wigglez} 
			& 1.48  & 0.462 & 0.045 & 2021 & \cite{hou2021eboss} \\
			
			0.51  & 0.516 & 0.062  & 2024 & \cite{Adame_2025} 
			& 1.49  & 0.435 & 0.045 & 2024 & \cite{Adame_2025} \\
			
			0.60  & 0.390 & 0.063  & 2012 & \cite{blake2012wigglez} 
			& 1.944 & 0.364 & 0.106 & 2019 & \cite{zhao2019_dr14_qso} \\
			\bottomrule[0.4mm]
		\end{tabular}
	\end{adjustbox}
\end{table}
\section{Constraints on Models}\label{Constraints-on-Models}

In this section, we employ the Markov Chain Monte Carlo (MCMC) technique to derive the best-fit values and confidence intervals of the cosmological parameters. Both the standard $\Lambda$CDM and the R-$\Lambda$CDM models are investigated under the assumptions of spatially flat and non-flat universes. 

To assess the robustness of the constraints and to quantify the impact of different cosmological probes, we consider four complementary combinations of observational data:
\begin{itemize}
	\item $\mathbf{D_1}$ = SN + $H(z)$ + $f\sigma_8(z)$,
	\item $\mathbf{D_2}$ = SN + $H(z)$ + $f\sigma_8(z)$ + BAO + BBN,
	\item $\mathbf{D_3}$ = SN + $H(z)$ + $f\sigma_8(z)$ + BAO + CMB,
	\item $\mathbf{D_4}$ = CMB (Planck 2018 distance priors only).
\end{itemize}

Here, SN denotes the Pantheon+SH0ES Type Ia supernova sample, $H(z)$ corresponds to the compilation of cosmic chronometer and BAO-derived Hubble measurements, $f\sigma_8(z)$ represents the growth-rate data, BAO refers to the DESI DR2 clustering measurements, BBN is implemented via the PRIMAT helium abundance constraint, and CMB corresponds to the Planck 2018 distance priors.

The dataset $\mathbf{D_1}$ contains exclusively late-time probes and therefore constrains the expansion history and growth of structure without imposing early-Universe information. The combination $\mathbf{D_2}$ supplements late-time data with primordial nucleosynthesis constraints, while $\mathbf{D_3}$ incorporates compressed CMB information from recombination. 

In addition, the dataset $\mathbf{D_4}$ is introduced to isolate the contribution of early-Universe information by considering the Planck 2018 CMB distance priors alone. This allows for a direct determination of the Hubble constant from CMB observations, independently of late-time probes, and provides a reference baseline for interpreting the Hubble tension.

This hierarchical dataset construction enables us to explicitly evaluate how different observational regimes affect the inferred cosmological parameters, particularly the Hubble constant $H_0$ and the amplitude of matter fluctuations $\sigma_8$.

The results of the parameter estimation are summarized in Tables~\ref{tab:results_flat_LCDM}--\ref{tabTre}, while the corresponding two-dimensional confidence contours are illustrated in Figures~\ref{fig:lcdm_flat_fig1}--\ref{fig:rlcdm_nonflat_fig1_k<0}.

\subsection*{Model selection criteria}

To compare the statistical performance of the models, we employ the Akaike Information Criterion (AIC) and the Bayesian Information Criterion (BIC), which are widely used tools in cosmological model selection \cite{capozziello2011comprehensive,kass1995bayes,spiegelhalter2002bayesian,whitehead2007selection,liddle2007information,rezaei2021comparison}. The AIC and BIC are defined as
\begin{equation}
	\mathrm{AIC} = \chi^2_{\min} + 2d,
\end{equation}
\begin{equation}
	\mathrm{BIC} = \chi^2_{\min} + d \ln N,
\end{equation}
where $d$ is the number of free cosmological parameters in the model and $N$ is the total number of data points included in a given dataset combination (accounting for all measurements entering the likelihood). For model comparison within each dataset combination, we consider the differences
\begin{equation}
	\Delta \mathrm{AIC} = \mathrm{AIC}_{\rm model} - \mathrm{AIC}_{\rm min},
	\quad
	\Delta \mathrm{BIC} = \mathrm{BIC}_{\rm model} - \mathrm{BIC}_{\rm min},
\end{equation}
where the reference value corresponds to the model with the minimum information criterion for that specific dataset. According to the $\Delta\mathrm{AIC}$ criterion,  $\Delta\mathrm{AIC} \leq 2$ indicates substantial support for a model, $4 \leq \Delta\mathrm{AIC} \leq 7$ suggests considerably less support, and $\Delta\mathrm{AIC} \geq 10$ implies that the model is strongly disfavored. Similarly, in the context of BIC, $\Delta\mathrm{BIC} \leq 2$ is regarded as weak evidence, $2 \leq \Delta\mathrm{BIC} \leq 6$ indicates positive evidence, $6 \leq \Delta\mathrm{BIC} \leq 10$ reflects strong evidence, and values exceeding 10 are interpreted as very strong evidence against the model with the higher BIC \cite{kass1995bayes}.

\section{Numerical results}\label{Results}

\subsection{Constraining the flat and non-flat models of $\Lambda$CDM}

In this subsection, we present the parameter constraints obtained for both the spatially flat and non-flat $\Lambda$CDM models using the dataset combinations defined in Section~\ref{Constraints-on-Models}. For each geometry, we derive the marginalized best-fit values and $1\sigma$ confidence intervals of the cosmological parameters, together with the minimum chi-square and the corresponding information criteria.

For the spatially flat model, the sampled parameter set consists of ($\Omega_{dm0}$,  $\Omega_{b0}$,  $H_0$, $\sigma_8$,  $M_B$), while the radiation density parameter is derived assuming fixed $N_{\rm eff}=3.046$. The dark-energy density parameter $\Omega_{\Lambda 0}$ is obtained from the Friedmann closure condition, $\Omega_{\Lambda 0} = 1 - \Omega_{m0} - \Omega_{r0}$, with $\Omega_{m0}=\Omega_{dm0}+\Omega_{b0}$. In the non-flat case, the curvature parameter $\Omega_{k0}$ is included as an additional free parameter, and the closure relation becomes
$\Omega_{\Lambda 0} = 1 - \Omega_{m0} - \Omega_{r0} - \Omega_{k0}$.

We first examine the constraints obtained from the late-time dataset $\mathbf{D_1}$, which provides a baseline determination of $H_0$ and $\sigma_8$ without imposing early-Universe priors. We then investigate how the inclusion of primordial information through BBN ($\mathbf{D_2}$) or CMB distance priors ($\mathbf{D_3}$) modifies the posterior distributions. Particular attention is devoted to the behavior of the Hubble constant, the curvature parameter, and the stability of $\sigma_8$ across datasets.

In addition, we include the dataset $\mathbf{D_4}$, corresponding to the Planck 2018 CMB distance priors alone, as a reference baseline. This configuration allows us to isolate the constraints driven purely by early-Universe information and to directly compare them with those obtained from late-time and combined datasets. The D$_4$ dataset is therefore used as an anchor to assess how the inclusion of different cosmological probes shifts the inferred value of $H_0$ relative to the Planck-preferred region.

To quantify the level of agreement or tension between different dataset combinations, we employ the statistical estimator described in Appendix~\ref{appendix:tension-estimator}. This allows us to assess whether shifts in parameter means between $\mathbf{D_1}$, $\mathbf{D_2}$, $\mathbf{D_3}$, and the Planck baseline $\mathbf{D_4}$ are statistically significant or remain within the expected combined uncertainties.

The complete numerical results are summarized in Tables~\ref{tab:results_flat_LCDM}--\ref{VII-CPD}, while the corresponding two-dimensional marginalized confidence contours are shown in Figures~\ref{fig:lcdm_flat_fig1}--\ref{fig:lcdm_nonflat_fig1_k<0}. The posterior distributions for the Planck-only D$_4$ dataset are presented separately in Fig.~\ref{fig:triangle_flat_LCDM_D4}, illustrating the tight constraints provided by early-Universe observations. These results provide a comprehensive assessment of the internal consistency of the $\Lambda$CDM framework under both flat and non-flat assumptions.

\subsubsection{Flat $\Lambda$CDM model}

We perform a comprehensive analysis of the dataset combinations D$_1$, D$_2$, and D$_3$ within the framework of the spatially flat $\Lambda$CDM cosmology. For reference, the Planck baseline (denoted here as D$_4$) is included only for external comparison. The resulting 68\% confidence level constraints on the cosmological parameters are summarized in Table~\ref{tab:results_flat_LCDM}.  In this analysis, the sampled parameters are $\Omega_{dm_0}$, $\Omega_{b_0}$, $H_0$, and $\sigma_8$, while $\Omega_{\Lambda_0}$ and $\Omega_{r_0}$ are derived from the flatness condition and the fixed radiation sector ($N_{\rm eff}=3.046$). The corresponding one- and two-dimensional marginalized constraints for D$_1$--D$_3$ are shown in Fig.~\ref{fig:lcdm_flat_fig1}.

The inferred Hubble constant spans the range $67.35 \leq H_0 \leq 71.30$ km\,s$^{-1}$\,Mpc$^{-1}$. In particular, dataset D$_2$ yields $H_0 = 71.04^{+0.72}_{-0.73}$ km\,s$^{-1}$\,Mpc$^{-1}$. This value differs from the Planck 2018 estimate ($67.4 \pm 0.5$) by $4.13\sigma$ and from the SH0ES measurement ($73.2 \pm 1.3$) by $1.45\sigma$. For comparison, the direct Planck--SH0ES discrepancy amounts to approximately $4.16\sigma$. Thus, late-time datasets shift the inferred expansion rate toward the local value, though without reconciling both measurements simultaneously.

\subsubsection*{Comparison of D$_1$ with D$_2$ and D$_3$}

Between D$_1$ and D$_2$, all parameters remain consistent within less than $1\sigma$. The deviations for $\Omega_{dm_0}$, $\Omega_{b_0}$, $H_0$, and $\sigma_8$ are $0.35\sigma$, $0.76\sigma$, $0.24\sigma$, and $0.08\sigma$, respectively. The derived parameters shift by $0.43\sigma$ ($\Omega_{\Lambda_0}$) and $0.24\sigma$ ($\Omega_{r_0}$).

In contrast, D$_3$ (including CMB priors) shows larger deviations relative to D$_1$: $1.18\sigma$ for $\Omega_{dm_0}$, $0.28\sigma$ for $\Omega_{b_0}$, $4.66\sigma$ for $H_0$, and $0.45\sigma$ for $\sigma_8$, with derived shifts of $2.44\sigma$ in $\Omega_{\Lambda_0}$ and $4.96\sigma$ in $\Omega_{r_0}$. These shifts primarily reflect the strong constraining power of CMB information on the background expansion, in particular on $H_0$, with corresponding shifts in derived quantities.

\subsubsection*{Comparison with the D$_4$ (Planck) baseline}

The dataset D$_4$, corresponding to the Planck 2018 CMB distance priors alone, provides a direct determination of cosmological parameters from early-Universe information and serves as a reference baseline for interpreting the Hubble tension within a consistent framework.

Using the constraints summarized in Table~\ref{tab:results_flat_LCDM}, we find that the late-time dataset D$_1$ exhibits a significant deviation from the Planck baseline, with a discrepancy of approximately $3.63\sigma$ in $H_0$ and $1.80\sigma$ in $\Omega_{\Lambda_0}$. A similar behavior is observed for D$_2$, with differences of $3.58\sigma$ in $H_0$ and $1.93\sigma$ in $\Omega_{\Lambda_0}$.

\begin{figure}[H]
	\centering
	\includegraphics[width=0.8\textwidth]{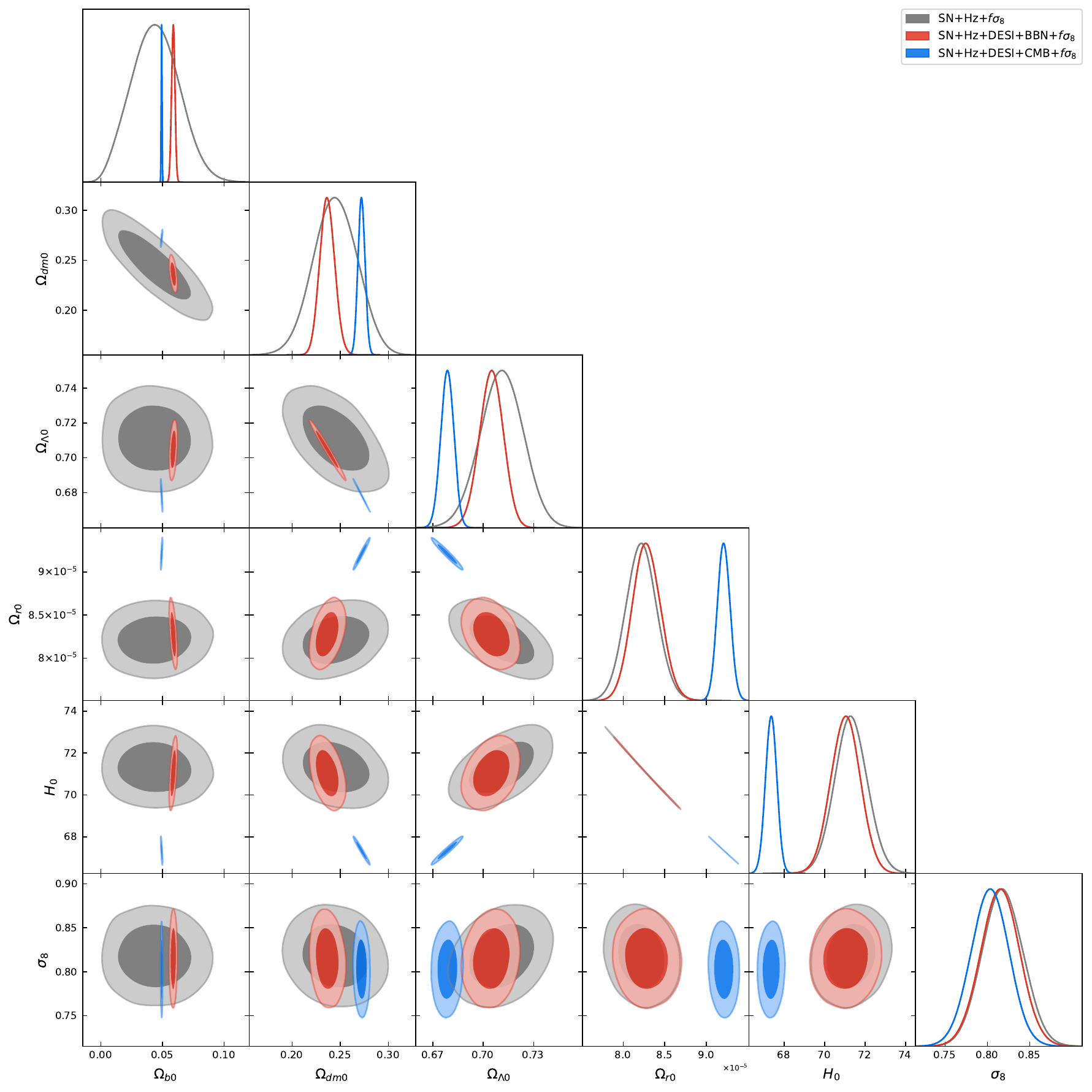}
	\caption{\small One-dimensional likelihoods and 1$\sigma$ and $2\sigma$ confidence contours for the flat $\Lambda$CDM parameters inferred from D$_1$ (SN+$H(z)$+$f\sigma_8$), D$_2$ (SN+$H(z)$+DESI+BBN+$f\sigma_8$), and D$_3$ (SN+$H(z)$+DESI+CMB priors+$f\sigma_8$).}
	\label{fig:lcdm_flat_fig1}
\end{figure}

In contrast, the CMB-inclusive dataset D$_3$ shows a much closer agreement with the Planck baseline, with deviations reduced to $0.47\sigma$ in $H_0$ and $0.53\sigma$ in $\Omega_{\Lambda_0}$. This behavior is also clearly reflected in the posterior distributions shown in Figures~\ref{fig:lcdm_flat_fig1} and \ref{fig:triangle_flat_LCDM_D4}, where the inclusion of CMB information leads to a significant tightening of the parameter constraints and shifts the inferred value of $H_0$ toward the Planck-preferred region.

These results demonstrate that the dataset D$_4$ provides a stable early-Universe anchor for the cosmological parameters, while late-time probes alone (D$_1$ and D$_2$) systematically favor higher values of $H_0$. The inclusion of CMB information in D$_3$ effectively bridges these regimes, highlighting that the observed variation in $H_0$ is primarily driven by the dataset composition rather than by intrinsic modifications of the background cosmological model.

\subsubsection*{AIC and BIC analysis for flat $\Lambda$CDM model}

For D$_1$, we obtain AIC = 1581.53 and BIC = 1608.92; for D$_2$, AIC = 1593.34 and BIC = 1620.76; and for D$_3$, AIC = 1642.93 and BIC = 1670.36. Since the number of free parameters is identical across datasets, these differences quantify the relative goodness-of-fit under different observational combinations rather than model preference. 

Overall, the flat $\Lambda$CDM analysis reveals a pronounced dependence of the inferred $H_0$ value on dataset selection, with late-time combinations systematically preferring higher expansion rates while CMB-inclusive constraints restore consistency with Planck. No dataset combination simultaneously reconciles the Planck and SH0ES determinations. Hence, within the minimal spatially flat framework, the Hubble tension remains statistically significant.

\begin{table}[h]
	\caption{\small The 68\% confidence limits of flat $\Lambda$CDM model cosmological parameters (CPs) from D$_1$, D$_2$, D$_3$, and D$_4$ datasets. H$_0$ has units of km\,s$^{-1}$\,Mpc$^{-1}$. Here $\Omega_{m_0} = \Omega_{dm_0} + \Omega_{b_0}$.}
	\centering
	\setlength{\tabcolsep}{9pt}
	\renewcommand{\arraystretch}{1.5}
		\begin{tabular}{lcccc}
			\hline\hline
			\textbf{CPs} & \textbf{D$_1$} & \textbf{D$_2$} & \textbf{D$_3$} & \textbf{D$_4$}  \\
			\hline
			$\Omega_{b_0}$ & $0.0440^{+0.0195}_{-0.0193}$ & $0.0588^{+0.0013}_{-0.0014}$ & $0.0494 \pm 0.0003$ & $0.0488^{+0.0007}_{-0.0007}$ \\
			$\Omega_{dm_0}$ & $0.2447^{+0.0229}_{-0.0225}$ & $0.2363^{+0.0078}_{-0.0075}$ & $0.2719 \pm 0.0036$   & $0.2675^{+0.0079}_{-0.0076}$ \\
			$\Omega_{\Lambda_0}$ & $0.7110^{+0.0126}_{-0.0128}$ & $0.7048^{+0.0070}_{-0.0072}$ & $0.6786^{+0.0038}_{-0.0039}$ & $0.6835^{+0.0083}_{-0.0086}$ \\
			$\Omega_{r_0}$ & $(8.2172^{+0.1855}_{-0.1842}) \times 10^{-5}$ & $(8.2776^{+0.1727}_{-0.1651}) \times 10^{-5}$ & $(9.2088^{+0.0772}_{-0.0754}) \times 10^{-5}$ & $(9.1236^{+0.1660}_{-0.1607})\times 10^{-5}$ \\
			$H_0$ & $71.30^{+0.81}_{-0.79}$ & $71.04^{+0.72}_{-0.73}$ & $67.35 \pm 0.28$ & $67.66^{+0.60}_{-0.61}$ \\
			$\sigma_8$ & $0.8177^{+0.0243}_{-0.0237}$ & $0.8150^{+0.0227}_{-0.0229}$ & $0.8030^{+0.0222}_{-0.0224}$ & $-$ \\
			\hline
			$\rm AIC$ & $1581.53132$ & $1593.33913$ & $1642.92754$                & $6.00022$ \\
			$\rm BIC$ & $1608.91651$ & $1620.76378$ & $1670.35780$                & $3.29606$ \\
			\hline\hline
	\end{tabular}
	\label{tab:results_flat_LCDM}
\end{table}

\begin{figure}[H]
	\centering
	\includegraphics[width=0.8\linewidth]{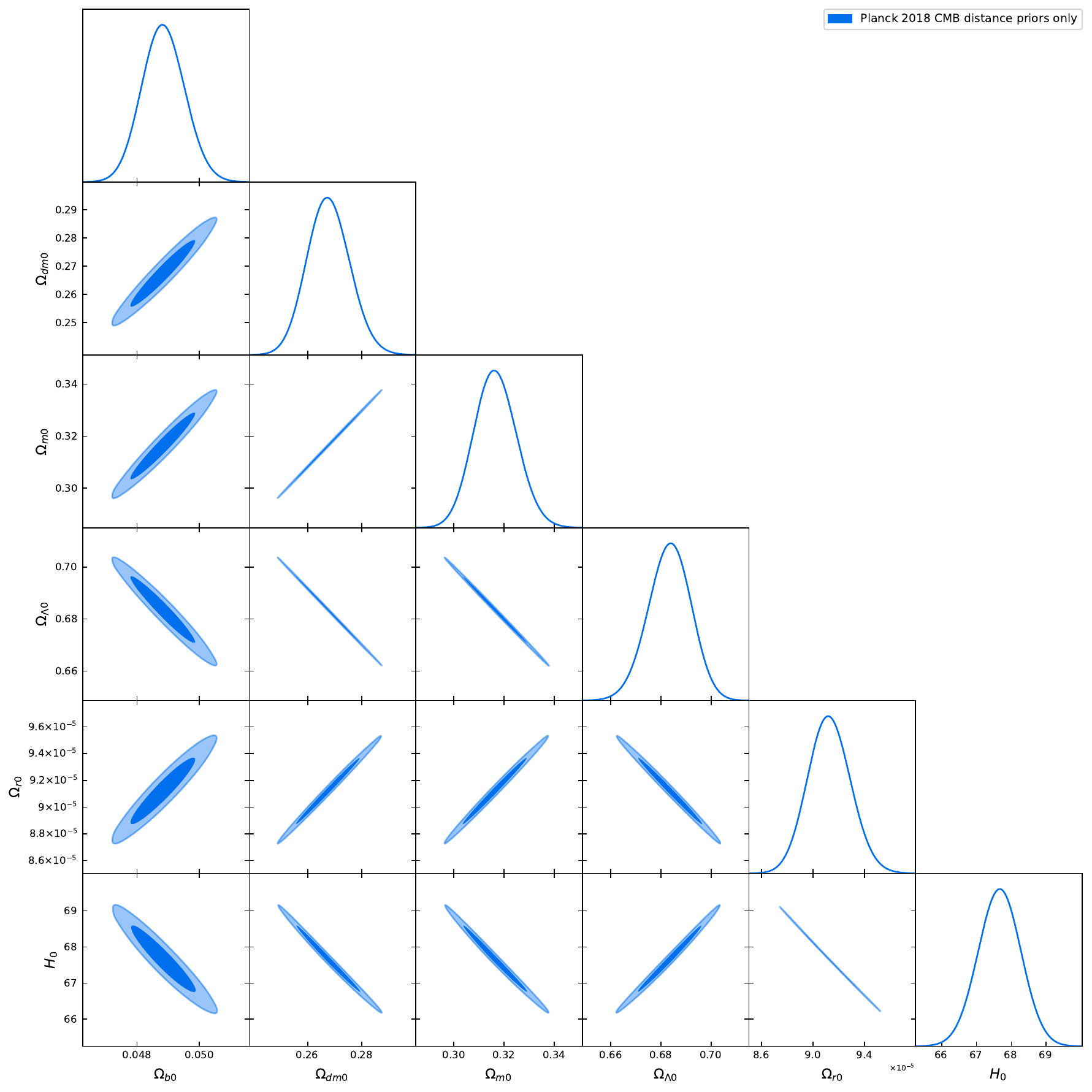}
	\caption{\small 
		Triangle plot of the posterior distributions for the flat $\Lambda$CDM model using the Planck 2018 CMB distance priors (D$_4$ dataset). The contours represent the $1\sigma$ and $2\sigma$ confidence levels. The results show tight constraints on all cosmological parameters, with $H_0$ consistent with the Planck baseline value.
	}
	\label{fig:triangle_flat_LCDM_D4}
\end{figure}
Figure~\ref{fig:triangle_flat_LCDM_D4} shows the posterior distributions of the cosmological parameters for the flat $\Lambda$CDM model using the Planck 2018 CMB distance priors (D$_4$ dataset). The contours are well localized and exhibit nearly Gaussian behavior, indicating that the compressed CMB likelihood provides tight and stable constraints on the standard cosmological parameters.

In particular, the Hubble constant is tightly constrained around $H_0 \sim 67$--$68~{\rm km\,s^{-1}\,Mpc^{-1}}$, in excellent agreement with the Planck baseline result. The strong degeneracy directions among parameters such as $\Omega_{m0}$ and $\Omega_{\Lambda0}$ follow the expected behavior of the standard $\Lambda$CDM framework, confirming the internal consistency of the dataset and the robustness of the inference.

\subsubsection{Open $\Lambda$CDM model}

\begin{figure}[H]
	\centering
	\mbox{\includegraphics[width=0.8\textwidth]{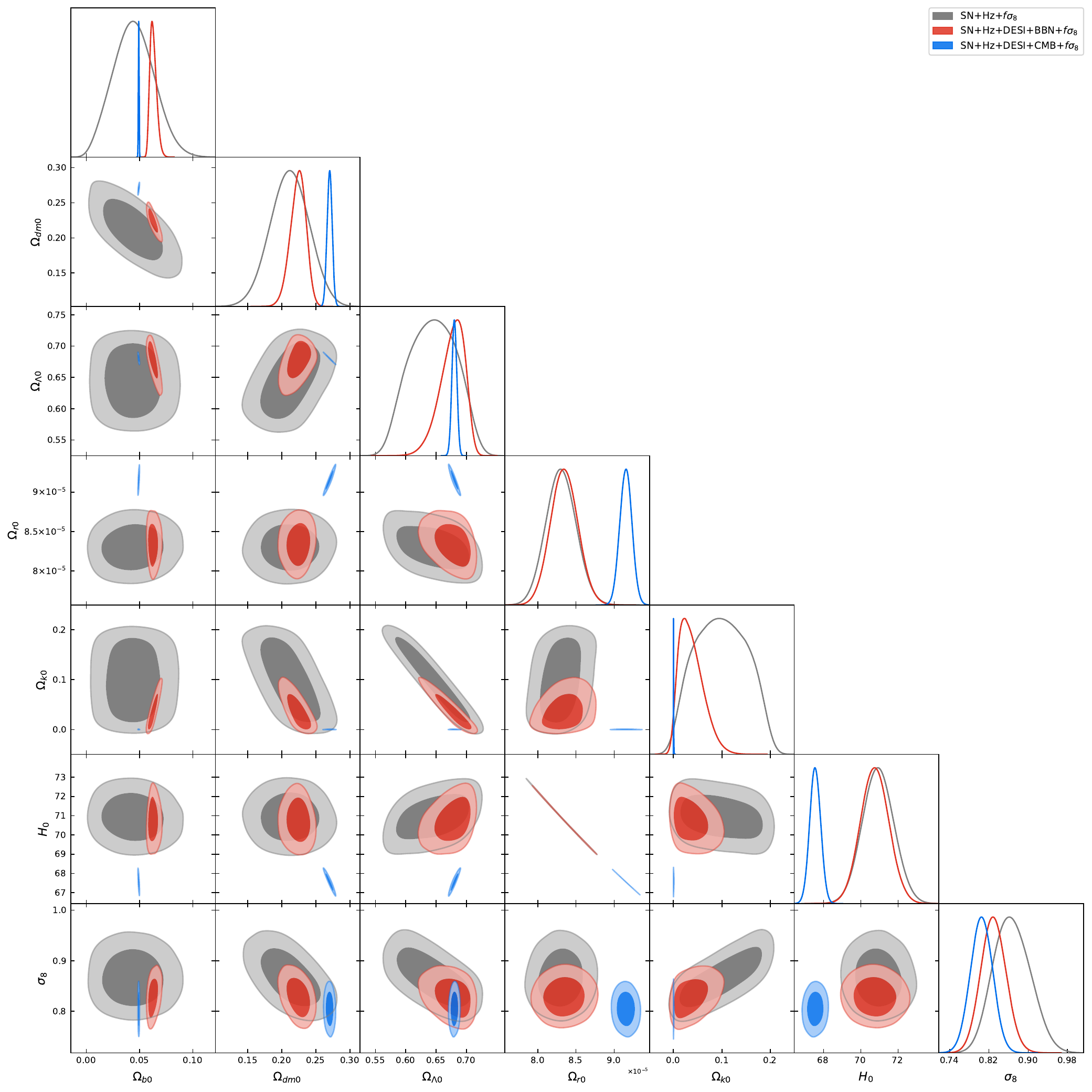}}
	\caption{\small One-dimensional likelihoods and 1$\sigma$ and $2\sigma$ confidence contours of open $\Lambda$CDM ($\Omega_{k_0}>0$) parameters inferred from D$_1$ (SN+$H(z)$+$f\sigma_8$), D$_2$ (SN+$H(z)$+DESI+BBN+$f\sigma_8$), and D$_3$ (SN+$H(z)$+DESI+CMB priors+$f\sigma_8$).}
	\label{fig:lcdm_nonflat_fig1_k>0}
\end{figure}

The open $\Lambda$CDM scenario extends the standard framework by introducing spatial curvature as an additional free parameter. The corresponding 68\% confidence level constraints obtained from D$_1$, D$_2$, and D$_3$ are summarized in Table~\ref{tab:results_open_LCDM}. The marginalized one- and two-dimensional likelihood contours for the three dataset combinations are shown in Fig.~\ref{fig:lcdm_nonflat_fig1_k>0}.
The inferred Hubble constant spans the range $67.55 \leq H_0 \leq 70.92$ km\,s$^{-1}$\,Mpc$^{-1}$. Relative to the Planck 2018 determination ($67.4 \pm 0.5$ km\,s$^{-1}$\,Mpc$^{-1}$), the deviations amount to $3.62\sigma$ (D$_1$), $3.72\sigma$ (D$_2$), and $0.26\sigma$ (D$_3$). With respect to the SH0ES local measurement ($73.2 \pm 1.3$ km\,s$^{-1}$\,Mpc$^{-1}$), the corresponding discrepancies are $1.48\sigma$, $1.63\sigma$, and $4.25\sigma$, respectively. Thus, as in the flat case, late-time dataset combinations favor systematically higher values of $H_0$, while the inclusion of CMB information restores agreement with the Planck baseline.

The curvature parameter $\Omega_{k_0}$ is positive for D$_1$ and D$_2$. In both cases the deviation from spatial flatness remains below the $2\sigma$ level. In contrast, D$_3$ yields $\Omega_{k_0}=0.0002^{+0.0004}_{-0.0002}$, which is fully consistent with a spatially flat Universe. This demonstrates the strong constraining power of CMB distance information on spatial curvature.

\subsubsection*{AIC and BIC analysis for open $\Lambda$CDM model}

The AIC and BIC values are 
(1582.21, 1615.07) for D$_1$, 
(1595.00, 1627.91) for D$_2$, 
and (1647.56, 1680.47) for D$_3$, 
as reported in Table~\ref{tab:results_open_LCDM}. 

Since the number of free parameters remains identical across the three dataset combinations, these criteria do not constitute model selection in the usual sense. Instead, they quantify the relative statistical performance of the open $\Lambda$CDM model under different observational inputs. 

Allowing for positive spatial curvature does not qualitatively alter the tension pattern observed in the flat case. The inferred curvature parameter remains statistically consistent with spatial flatness, and the shift in $H_0$ continues to depend primarily on the inclusion of CMB information. Consequently, introducing positive curvature does not provide a statistically meaningful resolution of the Hubble tension.
\begin{table}[h]
	\caption{\small The 68\% confidence limits of open $\Lambda$CDM model cosmological parameters (CPs) from D$_1$, D$_2$, and D$_3$ datasets. H$_0$ has units of km s$^{-1}$ Mpc$^{-1}$. Here $\Omega_{m_0} = \Omega_{dm_0} + \Omega_{b_0}$.}
	\centering
	\setlength{\tabcolsep}{9pt}
	\renewcommand{\arraystretch}{1.5}
		\begin{tabular}{lccc}
			\hline\hline
			\textbf{CPs} & \textbf{D$_1$} & \textbf{D$_2$} & \textbf{D$_3$} \\
			\hline
			$\Omega_{b_0}$ & $0.0441^{+0.0195}_{-0.0192}$ & $0.0625^{+0.0035}_{-0.0027}$ & $0.0492 \pm 0.0003$ \\
			$\Omega_{dm_0}$ & $0.2116^{+0.0284}_{-0.0285}$ & $0.2244^{+0.0106}_{-0.0122}$ & $0.2702^{+0.0036}_{-0.0037}$ \\
			$\Omega_{\Lambda_0}$ & $0.6455^{+0.0400}_{-0.0413}$ & $0.6789^{+0.0178}_{-0.0231}$ & $0.6802^{+0.0040}_{-0.0039}$ \\
			$\Omega_{r_0}$ & $(8.3036^{+0.1974}_{-0.1938}) \times 10^{-5}$ & $(8.3456^{+0.1788}_{-0.1749}) \times 10^{-5}$ & $(9.1536^{+0.0775}_{-0.0793}) \times 10^{-5}$ \\
			$\Omega_{k_0}$ & $0.0983^{+0.0591}_{-0.0578}$ & $0.0333^{+0.0289}_{-0.0218}$ & $0.0002^{+0.0004}_{-0.0002}$ \\
			$H_0$ & $70.92^{+0.84}_{-0.83}$ & $70.75^{+0.75}_{-0.75}$ & $67.55^{+0.29}_{-0.28}$ \\
			$\sigma_8$ & $0.8655^{+0.0395}_{-0.0361}$ & $0.8288^{+0.0259}_{-0.0250}$ & $0.8048 \pm 0.0224$ \\
			\hline
			$\rm AIC$ & $1582.21109$ & $1595.00379$ & $1647.55658$ \\
			$\rm BIC$ & $1615.07332$ & $1627.91337$ & $1680.47290$ \\
			\hline\hline
	\end{tabular}
	\label{tab:results_open_LCDM}
\end{table}

\subsubsection{Closed $\Lambda$CDM model}

\begin{figure}[H]
	\centering
	\mbox{\includegraphics[width=0.8\textwidth]{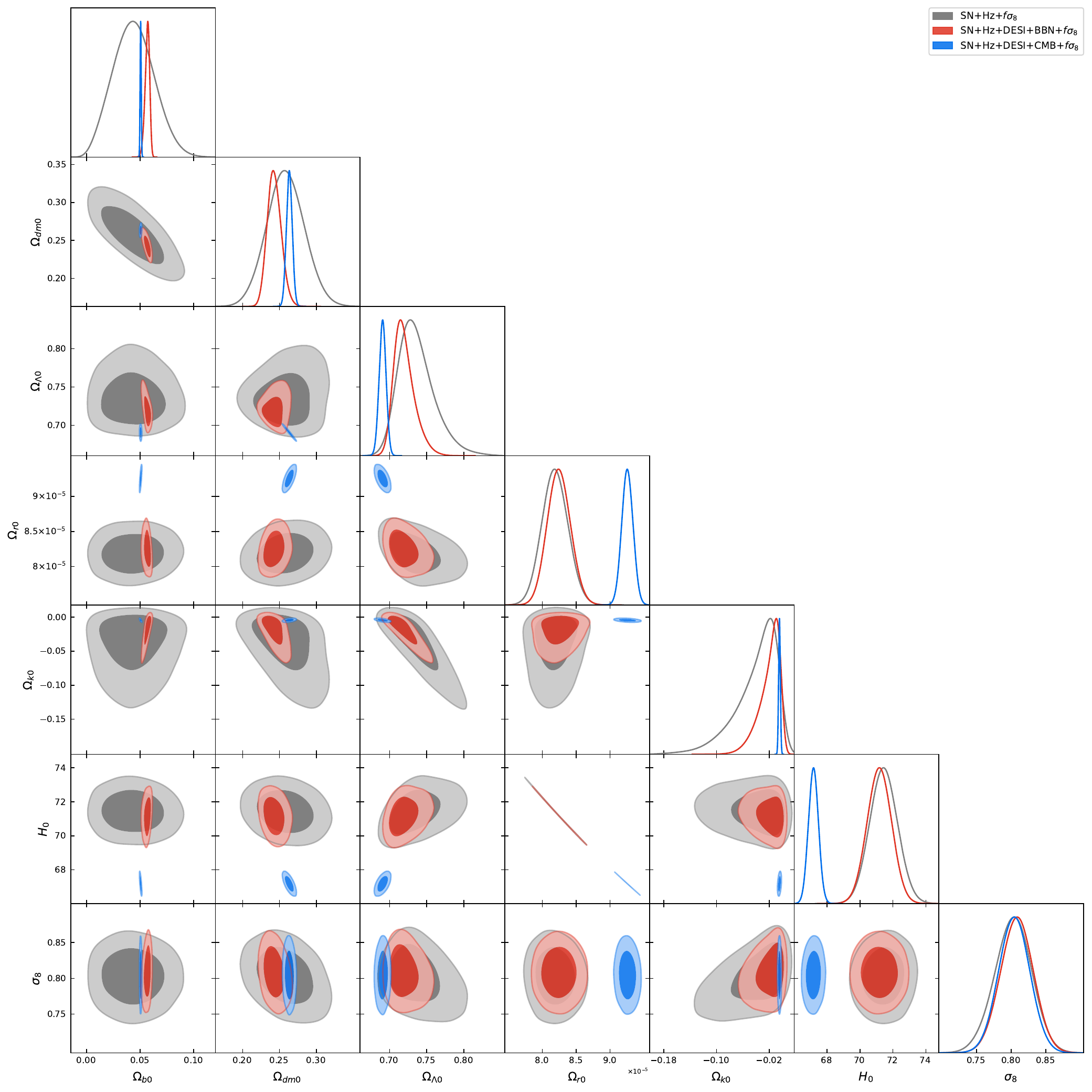}}
	\caption{\small One-dimensional marginalized likelihoods and 1$\sigma$ and $2\sigma$ confidence contours of the closed $\Lambda$CDM ($\Omega_{k_0}<0$) model parameters inferred from D$_1$ (SN+$H(z)$+$f\sigma_8$), D$_2$ (SN+$H(z)$+DESI+BBN+$f\sigma_8$), and D$_3$ (SN+$H(z)$+DESI+CMB priors+$f\sigma_8$).}
	\label{fig:lcdm_nonflat_fig1_k<0}
\end{figure}

Cosmological parameter constraints for the seven-parameter closed $\Lambda$CDM model, derived from the dataset combinations D$_1$, D$_2$, and D$_3$, are summarized in Table~\ref{tab:results_closed_LCDM}. The corresponding marginalized likelihood contours for the three dataset combinations are shown in Fig.~\ref{fig:lcdm_nonflat_fig1_k<0}.

The Hubble constant spans the range $67.18 \leq H_0 \leq 71.43$ km\,s$^{-1}$\,Mpc$^{-1}$. Relative to the Planck 2018 determination ($67.4 \pm 0.5$ km\,s$^{-1}$\,Mpc$^{-1}$), the deviations amount to $4.14\sigma$, $4.23\sigma$, and $0.38\sigma$ for D$_1$, D$_2$, and D$_3$, respectively. Thus, only D$_3$ remains statistically consistent with Planck.

With respect to the SH0ES local measurement ($73.2 \pm 1.3$ km\,s$^{-1}$\,Mpc$^{-1}$), the discrepancies are $1.15\sigma$ (D$_1$), $1.35\sigma$ (D$_2$), and $4.51\sigma$ (D$_3$). As in the flat and open cases, the late-time combinations D$_1$ and D$_2$ prefer higher values of $H_0$ closer to the local determination, whereas the inclusion of CMB information in D$_3$ restores agreement with Planck while maintaining a substantial discrepancy with SH0ES. Therefore, the overall Planck–SH0ES tension is not resolved within the closed $\Lambda$CDM framework.

The curvature parameter $\Omega_{k_0}$ is negative in all datasets, corresponding to a closed spatial geometry. For D$_1$ and D$_2$, the inferred values ($-0.0306^{+0.0221}_{-0.0393}$ and $-0.0156^{+0.0113}_{-0.0199}$) remain statistically consistent with spatial flatness at approximately the $1\sigma$ level. 
In contrast, D$_3$ yields $\Omega_{k_0} = -0.0047 \pm 0.0013$, corresponding to a statistically significant $\sim3.6\sigma$ deviation from zero. This indicates a statistically significant preference for negative curvature when CMB priors are included.

The dark matter density parameter $\Omega_{dm_0}$ varies within the interval $0.243 \lesssim \Omega_{dm_0} \lesssim 0.263$ across the three dataset combinations, while the baryon density parameter $\Omega_{b_0}$ lies in the range $0.044$–$0.057$. The structure growth parameter $\sigma_8$ remains close to the $\sim0.80$ level for the late-time combinations D$_1$ and D$_2$, whereas the inclusion of CMB priors in D$_3$ shifts the preferred value to $\sigma_8 \simeq 0.87$. This increase reflects the strong correlation between spatial curvature and the growth amplitude in the closed geometry, highlighting the impact of CMB distance information on both background and perturbation parameters.

\subsubsection*{AIC and BIC analysis for closed $\Lambda$CDM model}

The AIC and BIC values for the closed $\Lambda$CDM model are reported in Table~\ref{tab:results_closed_LCDM}. For D$_1$, we obtain AIC = 1583.72 and BIC = 1616.58; for D$_2$, AIC = 1595.45 and BIC = 1628.36; and for D$_3$, AIC = 1634.51 and BIC = 1667.42.

Since the number of free parameters remains identical across D$_1$, D$_2$, and D$_3$, 
these criteria quantify the relative goodness-of-fit of the same cosmological model under different observational combinations rather than indicating intrinsic model preference.

In the closed $\Lambda$CDM scenario, CMB priors indicate a statistically significant ($\sim3.6\sigma$) preference for negative curvature; however, this geometric extension does not enable a simultaneous agreement between early- and late-universe measurements of $H_0$. As in the flat and open geometries, late-time datasets favor higher expansion rates, whereas CMB-inclusive constraints align closely with Planck. Therefore, the persistence of the Hubble tension appears to be driven primarily by dataset dependence rather than by spatial curvature.

\begin{table}[h]
	\caption{\small The 68\% confidence limits of closed $\Lambda$CDM model cosmological parameters (CPs) from D$_1$, D$_2$, and D$_3$ datasets. H$_0$ has units of km s$^{-1}$ Mpc$^{-1}$. Here $\Omega_{m_0} = \Omega_{dm_0} + \Omega_{b_0}$.}
	\centering
	\setlength{\tabcolsep}{9pt}
	\renewcommand{\arraystretch}{1.5}
		\begin{tabular}{lccc}
			\hline\hline
			\textbf{CPs} & \textbf{D$_1$} & \textbf{D$_2$} & \textbf{D$_3$} \\
			\hline
			$\Omega_{b_0}$ & $0.0436^{+0.0194}_{-0.0190}$ & $0.0569^{+0.0019}_{-0.0023}$ & $0.0505 \pm 0.0005$ \\
			$\Omega_{dm_0}$ & $0.2579^{+0.0256}_{-0.0247}$ & $0.2426^{+0.0099}_{-0.0089}$ & $0.2634^{+0.0040}_{-0.0039}$ \\
			$\Omega_{\Lambda_0}$ & $0.7332^{+0.0266}_{-0.0202}$ & $0.7179^{+0.0156}_{-0.0112}$ & $0.6907 \pm 0.0046$ \\
			$\Omega_{r_0}$ & $(8.1853^{+0.1941}_{-0.1886}) \times 10^{-5}$ & $(8.2444^{+0.1729}_{-0.1677}) \times 10^{-5}$ & $(9.2550 \pm 0.0814) \times 10^{-5}$ \\
			$\Omega_{k_0}$ & $-0.0306^{+0.0221}_{-0.0393}$ & $-0.0156^{+0.0113}_{-0.0199}$ & $-0.0047 \pm 0.0013$ \\
			$H_0$ & $71.43^{+0.84}_{-0.83}$ & $71.18 \pm 0.74$ & $67.18 \pm 0.30$ \\
			$\sigma_8$ & $0.8032^{+0.0257}_{-0.0268}$ & $0.8084^{+0.0232}_{-0.0236}$ & $0.8045^{+0.0222}_{-0.0223}$ \\
			\hline
			$\rm AIC$ & $1583.71522$ & $1595.45415$ & $1634.50770$ \\
			$\rm BIC$ & $1616.57745$ & $1628.36373$ & $1667.42402$ \\
			\hline\hline
	\end{tabular}
	\label{tab:results_closed_LCDM}
\end{table}

\subsubsection{Assessment of geometrical models based on parameter consistency: The $\Lambda$CDM model}

A comparative examination of the cosmological parameter differences between datasets D$_1$, D$_2$, and D$_3$, summarized in Table~\ref{VII-CPD}, enables us to assess the internal consistency of the flat, open, and closed $\Lambda$CDM geometries. The deviations are expressed in units of the combined statistical uncertainty ($\sigma$) between D$_1$ and the other dataset combinations, computed using symmetrized errors. This analysis is independent of external reference values (such as Planck or SH0ES) and instead provides a direct measure of the internal consistency among the dataset combinations themselves.

A clear pattern emerges across all three geometrical scenarios. For the D$_1$–D$_2$ comparison, the majority of parameters exhibit deviations at or below the $1\sigma$ level in flat, open, and closed models. This indicates strong mutual consistency among late-time datasets (SN, $H(z)$, $f\sigma_8$, BAO, and BBN), largely independent of the assumed spatial curvature.

In contrast, the comparison between D$_1$ and D$_3$ reveals substantially larger shifts in several parameters, most notably in $H_0$ and $\Omega_{r_0}$. In the flat model, the deviation in $H_0$ reaches $4.6\sigma$, while in the open and closed models it amounts to $3.8\sigma$ and $4.8\sigma$, respectively. A similar pattern is observed for $\Omega_{r_0}$, where deviations approach the $4$–$5\sigma$ level across geometries. It is important to note that $\Omega_{r_0}$ is not independently sampled in our analysis but is derived assuming a fixed effective number of relativistic species, $N_{\rm eff}=3.046$. Therefore, the apparent large shift in $\Omega_{r_0}$ primarily reflects its dependence on $H_0$ rather than any modification of the underlying radiation physics. These large shifts arise from the inclusion of CMB distance priors in D$_3$, which strongly constrain the background geometry and expansion history.

The curvature parameter $\Omega_{k_0}$ itself shows only moderate deviations (below $2\sigma$) when comparing D$_1$ with D$_3$ in the non-flat cases. This suggests that spatial curvature is not the primary driver of parameter inconsistencies between dataset combinations. Rather, the dominant effect originates from the addition of CMB information, which pulls all geometrical scenarios toward values consistent with Planck.

The structure growth parameter $\sigma_8$ remains relatively stable across dataset combinations, particularly in the closed case where deviations are well below $1\sigma$. This further supports the conclusion that late-time probes are mutually consistent and that the principal shifts arise when early-universe information is incorporated.

In the closed $\Lambda$CDM scenario, CMB priors indicate a mild preference for negative curvature; however, this geometric extension does not enable a simultaneous agreement between early- and late-universe measurements of $H_0$. As in the flat and open geometries, late-time datasets favor higher expansion rates, whereas CMB-inclusive constraints align closely with Planck. Therefore, the persistence of the Hubble tension appears to be driven by dataset dependence rather than by spatial curvature.
\begin{table}[h]
	\caption{\small Cosmological Parameter Differences (CPD) expressed in units of $\sigma$ for the flat, open, and closed $\Lambda$CDM models. The deviations are computed between D$_1$ and D$_2$, and between D$_1$ and D$_3$, using symmetrized uncertainties. The Planck-only dataset D$_4$ is treated separately as an early-Universe reference baseline. Values are rounded to one decimal place.}
	\centering
	\setlength{\tabcolsep}{3pt}
	\renewcommand{\arraystretch}{1.1}
		\begin{tabular}{l
				cc @{\hskip 10pt}
				cc @{\hskip 10pt}
				cc}
			\hline\hline
			\textbf{CPD}
			& \multicolumn{2}{c}{\textbf{$\Lambda$CDM (Flat)}}
			& \multicolumn{2}{c}{\textbf{$\Lambda$CDM (Open)}}
			& \multicolumn{2}{c}{\textbf{$\Lambda$CDM (Closed)}} \\
			
			& $|\textbf{D$_1$}-\textbf{D$_2$}|$
			& $|\textbf{D$_1$}-\textbf{D$_3$}|$
			& $|\textbf{D$_1$}-\textbf{D$_2$}|$
			& $|\textbf{D$_1$}-\textbf{D$_3$}|$
			& $|\textbf{D$_1$}-\textbf{D$_2$}|$
			& $|\textbf{D$_1$}-\textbf{D$_3$}|$ \\
			\hline
			
			$\Omega_{dm_0}$   & 0.4$\sigma$ & 1.2$\sigma$ & 0.4$\sigma$ & 2.0$\sigma$ & 0.5$\sigma$ & 0.2$\sigma$ \\
			
			$\Omega_{b_0}$    & 0.8$\sigma$ & 0.3$\sigma$ & 0.9$\sigma$ & 0.3$\sigma$ & 0.7$\sigma$ & 0.4$\sigma$ \\
			
			$\Omega_{k_0}$    & --           & --           & 1.0$\sigma$ & 1.7$\sigma$ & 0.4$\sigma$ & 0.8$\sigma$ \\
			
			$\Omega_{\Lambda_0}$ & 0.4$\sigma$ & 2.4$\sigma$ & 0.7$\sigma$ & 0.9$\sigma$ & 0.6$\sigma$ & 1.7$\sigma$ \\
			
			$\Omega_{r_0}$    & 0.2$\sigma$ & 5.0$\sigma$ & 0.2$\sigma$ & 4.0$\sigma$ & 0.2$\sigma$ & 4.8$\sigma$ \\
			
			$H_0$             & 0.2$\sigma$ & 4.6$\sigma$ & 0.2$\sigma$ & 3.8$\sigma$ & 0.2$\sigma$ & 4.8$\sigma$ \\
			
			$\sigma_8$        & 0.1$\sigma$ & 0.5$\sigma$ & 0.7$\sigma$ & 1.4$\sigma$ & 0.2$\sigma$ & 0.0$\sigma$ \\
			
			\hline\hline
	\end{tabular}
	\label{VII-CPD}
\end{table}
\subsection{Constraining the flat and non-flat models of R-$\Lambda$CDM}

In this subsection, we investigate the R-$\Lambda$CDM model as a minimal extension of the standard cosmological scenario, characterized by the inclusion of the Rastall parameter $\epsilon$. We consider spatially flat, open, and closed geometries, and derive observational constraints using the 
four dataset combinations defined in Section~\ref{Constraints-on-Models}.

For the spatially flat case, the sampled parameter set is given by ($\Omega_{dm0}$, $\Omega_{b0}$, $H_0$, $\sigma_8$, $\epsilon$, $M_B$), while the radiation density parameter is derived by fixing $N_{\rm eff}=3.046$. The effective density components entering the modified Friedmann equation are consistently constructed up to $\mathcal{O}(\epsilon^2)$, and the dark-energy density parameter is obtained from the generalized closure relation.

In the non-flat scenario, the curvature parameter $\Omega_{k0}$ is included as an additional free parameter. This allows us to explore potential degeneracies between spatial curvature and the Rastall parameter, thereby providing a direct assessment of the interplay between geometric effects and modified-gravity contributions in determining the expansion history.

We begin by analyzing the constraints obtained from the late-time dataset $\mathbf{D_1}$, which isolates the impact of modified background dynamics in the absence of early-Universe priors. We then examine the effect of incorporating primordial information through Big Bang Nucleosynthesis ($\mathbf{D_2}$), as well as the combined constraints from baryon acoustic oscillations and CMB distance priors ($\mathbf{D_3}$). 
Finally, we consider the CMB-only dataset ($\mathbf{D_4}$), which serves as a direct Planck baseline. This dataset plays a crucial role in assessing the behavior of the Rastall parameter $\epsilon$ and its degeneracy with spatial curvature, independently of late-time observational inputs.

To quantify the level of agreement or tension among different dataset combinations, we employ the statistical estimator described in Appendix~\ref{appendix:tension-estimator}. 
In particular, the inclusion of $\mathbf{D_4}$ enables a direct comparison with the Planck baseline, allowing us to assess the consistency in the determination of $H_0$ between early- and late-Universe datasets.

The numerical results are summarized in Tables~\ref{tab:results_flat_RLCDM}--\ref{tabTre}, while the corresponding marginalized one- and two-dimensional confidence contours are presented in Figures~\ref{fig:rlcdm_flat_fig1}--\ref{fig:rlcdm_nonflat_fig1_k<0}. These results provide a systematic comparison between flat and curved geometries within the Rastall framework 
and explicitly highlight the role of the $\mathbf{D_4}$ dataset in establishing the Planck reference constraints, allowing us to assess the extent to which the inclusion of $\epsilon$ alleviates existing cosmological tensions.

\subsubsection{R-$\Lambda$CDM model in a flat universe}

\begin{figure}[H]
	\centering
	\includegraphics[width=0.8\linewidth]{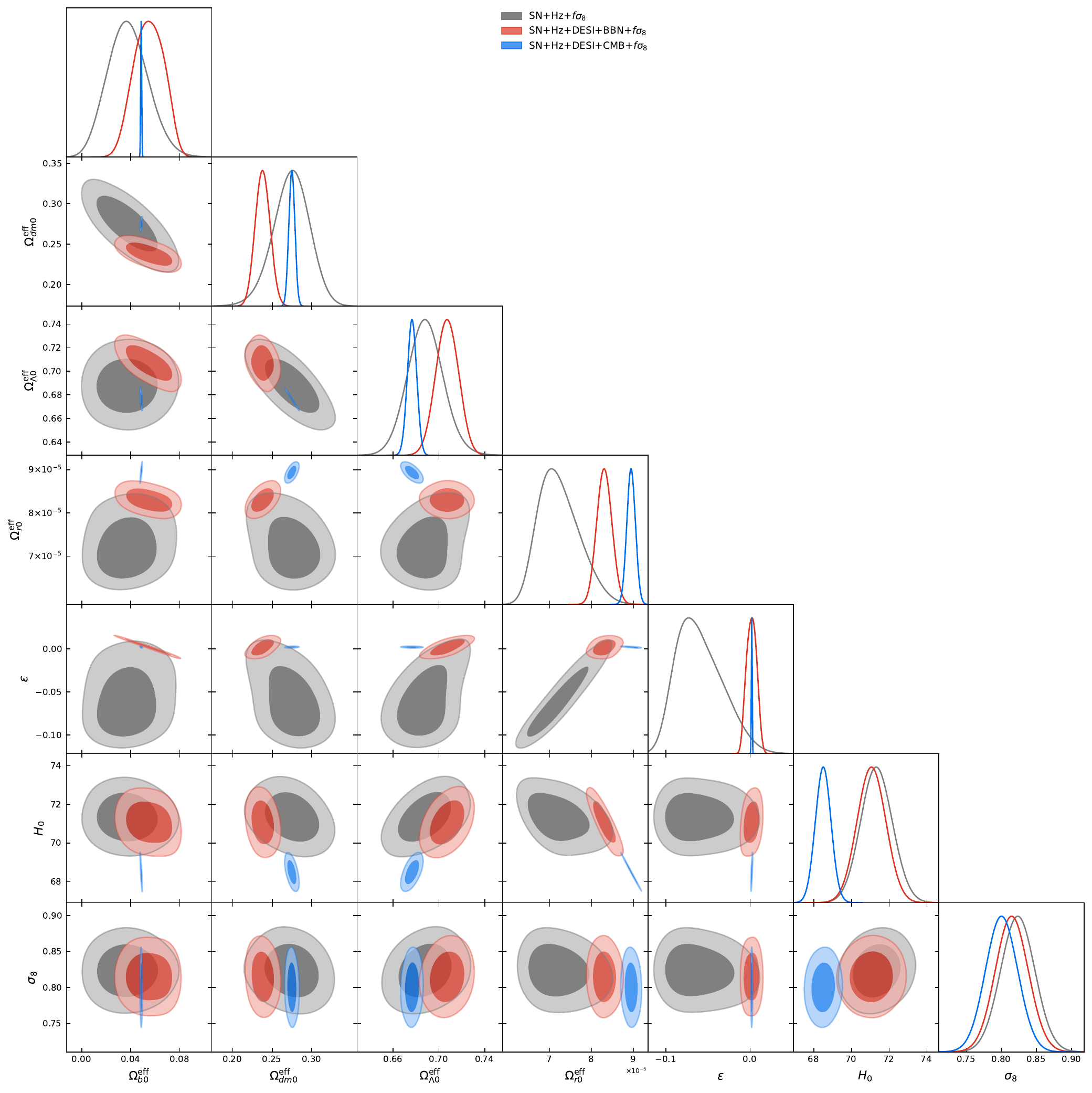}
	\caption{\small One-dimensional marginalized likelihoods and $1\sigma$ and $2\sigma$ confidence contours of the flat R-$\Lambda$CDM model parameters inferred from D$_1$ (SN+$H(z)$+$f\sigma_8$), D$_2$ (SN+$H(z)$+DESI+BBN+$f\sigma_8$), and D$_3$ (SN+$H(z)$+DESI+CMB priors+$f\sigma_8$). The displayed parameters include the effective density parameters $\Omega_{b0}^{\rm eff}$, $\Omega_{dm0}^{\rm eff}$, $\Omega_{\Lambda0}^{\rm eff}$, and $\Omega_{r0}^{\rm eff}$, together with the Rastall deviation parameter $\epsilon$, the Hubble constant $H_0$, and the clustering amplitude $\sigma_8$.}
	\label{fig:rlcdm_flat_fig1}
\end{figure}

In the flat R-$\Lambda$CDM framework, the deviation parameter $\epsilon$, commonly referred to as the Rastall parameter, introduces a modification to the standard $\Lambda$CDM dynamics. This modification originates from the non-conservation of the matter energy-momentum tensor due to a non-minimal coupling between matter and geometry (for further details, see subsection~\ref{subsec:RastalModel}). The results derived from this model, utilizing three distinct datasets D$_1$, D$_2$, and D$_3$, are summarized in Table~\ref{tab:results_flat_RLCDM} and depicted in Figure~\ref{fig:rlcdm_flat_fig1}.

The Hubble constant, $H_0$, is estimated to be approximately $71$ km s$^{-1}$ Mpc$^{-1}$ for datasets D$_1$ and D$_2$, while D$_3$ yields a lower value close to $68.5$ km s$^{-1}$ Mpc$^{-1}$. The discrepancies between these estimates and those obtained from Planck and the local SH0ES measurements show a clear dataset dependence. Relative to the Planck value, D$_1$ and D$_2$ remain in tension at roughly the $4\sigma$ level, whereas D$_3$ reduces this discrepancy to about $1.7\sigma$. Conversely, with respect to the local SH0ES determination, D$_1$ and D$_2$ are consistent within approximately $1.3\sigma$, while D$_3$ exhibits a larger deviation of about $3.5\sigma$. These results indicate that the flat R-$\Lambda$CDM scenario shifts the preferred value of $H_0$ depending on the adopted data combination, improving agreement with Planck when CMB information is included, but remaining closer to the local determination when only late-time data are considered.

The effective dark matter density parameter, $\Omega_{dm0}^{\rm eff}$, takes nearly identical central values in datasets D$_1$ and D$_3$, while D$_3$ substantially tightens the associated uncertainty. In contrast, D$_2$ prefers a lower value, indicating a noticeable shift in the inferred effective matter sector when the data combination is modified.
Similarly, the effective baryonic matter density parameter, $\Omega_{b0}^{\rm eff}$, shows a strong dependence on the inclusion of CMB information. While D$_1$ and D$_2$ allow comparatively broad ranges, D$_3$ yields a very tightly constrained value. The effective dark energy density parameter, $\Omega_{\Lambda0}^{\rm eff}$, remains within a narrow interval across all datasets, exhibiting moderate shifts between D$_2$ and the other combinations. The effective radiation density parameter, $\Omega_{r0}^{\rm eff}$, remains small in all cases, at the expected order of $10^{-5}$, although its precision improves once additional high-redshift information is incorporated.

The Rastall deviation parameter, $\epsilon$, displays a clear dataset dependence. Dataset D$_1$ favors a negative value, whereas D$_2$ and D$_3$ yield values fully consistent with zero, with D$_3$ providing significantly tighter constraints due to the inclusion of CMB information. This trend suggests that the departure from the standard $\Lambda$CDM limit becomes progressively suppressed as tighter observational constraints, particularly from CMB data, are included.

The parameter $\sigma_8$, which characterizes the amplitude of matter fluctuations, is constrained within the interval $0.80$–$0.82$ across the three datasets, with D$_3$ yielding slightly lower central values and tighter uncertainties. These values remain broadly compatible with CMB-based estimates within the quoted uncertainties.

Finally, information criteria, namely AIC and BIC, indicate that the dataset combination D$_1$ yields the lowest values among the three cases considered, while D$_2$ and D$_3$ result in comparatively larger values for their respective data combinations.

\subsubsection*{Comparison of dataset D$_1$ with D$_2$ and D$_3$}

The comparison of constrained parameters derived from the D$_1$ and D$_2$ datasets in the flat R-$\Lambda$CDM model reveals noticeable differences across several effective parameters. For example, the effective dark matter density parameter, $\Omega_{dm0}^{\rm eff}$, shows a shift of approximately $1.53\sigma$, indicating a lower value in D$_2$. Similarly, the effective radiation density $\Omega_{r0}^{\rm eff}$ and the effective cosmological constant $\Omega_{\Lambda0}^{\rm eff}$ exhibit deviations of approximately $2.45\sigma$ and $1.06\sigma$, respectively. The parameter $\epsilon$ shows a deviation of about $2.26\sigma$, reflecting a significant dataset dependence. In contrast, the Hubble constant $H_0$ remains highly consistent across the two datasets, differing only at the $\sim0.23\sigma$ level. These discrepancies, illustrated in Table~\ref{tabTre}, reflect the sensitivity of the effective parameter constraints to the adopted data combination.

The cosmological parameters derived from the D$_1$ and D$_3$ datasets show partial consistency but also reveal notable shifts in specific sectors. For instance, $H_0$ differs by approximately $3.07\sigma$. The effective radiation density $\Omega_{r0}^{\rm eff}$ and the Rastall parameter $\epsilon$ exhibit deviations of about $4\sigma$ and $2.32\sigma$, respectively. Meanwhile, $\Omega_{dm0}^{\rm eff}$ remains consistent at the central-value level, though D$_3$ significantly tightens its uncertainty. As summarized in Table~\ref{tabTre}, these results demonstrate that certain effective parameters are strongly affected by the inclusion of additional datasets.

\subsubsection*{Comparison with the D$_4$ (Planck) baseline}

The dataset D$_4$, based on the Planck 2018 CMB distance priors alone, provides a direct early-Universe reference for the flat R-$\Lambda$CDM model. As shown in Table~\ref{tab:results_flat_RLCDM}, this dataset yields $H_0 = 67.13^{+2.28}_{-1.04}~{\rm km\,s^{-1}\,Mpc^{-1}}$ and $\epsilon = 0.0023^{+0.0048}_{-0.0065}$, indicating that the Rastall parameter is consistent with the general relativity limit $\epsilon \simeq 0$, albeit with relatively large uncertainties due to parameter degeneracies in the compressed CMB likelihood.

In comparison, the late-time datasets D$_1$ and D$_2$ prefer higher values of the Hubble constant, $H_0 \sim 71~{\rm km\,s^{-1}\,Mpc^{-1}}$, with D$_1$ additionally showing a mild preference for negative values of $\epsilon$, while D$_2$ remains fully consistent with $\epsilon = 0$. These results reflect the increased flexibility of the model when only late-time information is considered.

A qualitatively different behavior emerges for the CMB-inclusive dataset D$_3$, which yields $H_0 = 68.50^{+0.40}_{-0.41}~{\rm km\,s^{-1}\,Mpc^{-1}}$ and tightly constrains the Rastall parameter to $\epsilon = 0.00203^{+0.00048}_{-0.00049}$. In this case, the inclusion of CMB information not only reduces the uncertainty in $H_0$, but also selects a small positive value of $\epsilon$ at high significance.

This comparison highlights the central role of dataset composition in the R-$\Lambda$CDM framework. While the Planck-only dataset D$_4$ anchors the model close to the $\Lambda$CDM limit with $\epsilon \approx 0$, it is the combined dataset D$_3$ that breaks parameter degeneracies and reveals a preference for small positive values of $\epsilon$. This, in turn, induces a mild but systematic upward shift in $H_0$ relative to the standard $\Lambda$CDM case, providing a direct observational manifestation of the $\epsilon$--$H_0$ degeneracy at the level of the background expansion.

Overall, these results demonstrate that the effect of the Rastall parameter on the inferred value of the Hubble constant is not intrinsic to the model alone, but emerges from the interplay between early- and late-Universe datasets. In particular, the enhancement of $H_0$ becomes visible only when $\epsilon$ is constrained to small positive values by CMB-inclusive data, while it remains suppressed when $\epsilon$ is weakly constrained or consistent with zero.

\begin{table}[h]
	\caption{\small The 68\% confidence limits of flat R-$\Lambda$CDM cosmological parameters from D$_1$, D$_2$, D$_3$, and D$_4$ datasets. $H_0$ is in km\,s$^{-1}$\,Mpc$^{-1}$. In Rastall gravity we report the \emph{effective} density parameters $\Omega_i^{\rm eff}\equiv\tilde{\Omega}_i$ that govern the background expansion.}
	\centering
	\setlength{\tabcolsep}{9pt}
	\renewcommand{\arraystretch}{1.5}
		\begin{tabular}{lcccc}
			\hline\hline
			\textbf{CPs} & \textbf{D$_1$} & \textbf{D$_2$} & \textbf{D$_3$} & \textbf{D$_4$} \\
			\hline
			$\Omega_{b0}^{\rm eff}$ & $0.03708^{+0.01676}_{-0.01600}$ & $0.05494^{+0.01326}_{-0.01278}$ & $0.04856 \pm 0.00039$ & $0.05018^{+0.00191}_{-0.00429}$  \\
			$\Omega_{dm0}^{\rm eff}$ & $0.2749^{+0.0216}_{-0.0232}$ & $0.2378^{+0.0094}_{-0.0092}$ & $0.2749^{+0.0037}_{-0.0038}$ & $0.29934^{+0.08302}_{-0.07910}$ \\
			$\Omega_{\Lambda0}^{\rm eff}$ & $0.6880^{+0.0153}_{-0.0150}$ & $0.7070^{+0.0099}_{-0.0097}$ & $0.6765^{+0.0040}_{-0.0039}$ & $0.65038^{+0.08340}_{-0.08507}$  \\
			$\Omega_{r0}^{\rm eff}$ & $(7.16 \pm 0.43)\times 10^{-5}$ & $(8.30 \pm 0.18)\times 10^{-5}$ & $(8.94 \pm 0.10)\times 10^{-5}$ & $(9.33^{+0.35}_{-0.74})\times 10^{-5}$  \\
			$\epsilon$ & $-0.06437^{+0.03287}_{-0.02447}$ & $0.00189^{+0.00615}_{-0.00662}$ & $0.00203^{+0.00048}_{-0.00049}$ & $0.002265^{+0.004823}_{-0.006460}$ \ \\
			$H_0$ & $71.32^{+0.84}_{-0.81}$ & $71.06 \pm 0.75$ & $68.50^{+0.40}_{-0.41}$ & $67.130^{+2.282}_{-1.041}$  \\
			$\sigma_8$ & $0.8232 \pm 0.0240$ & $0.8147^{+0.0232}_{-0.0228}$ & $0.8002^{+0.0226}_{-0.0225}$ & $-$ \\
			\hline
			$\mathrm{AIC}$ & $1580.01154$ & $1595.12210$ & $1628.04364$ & $8.00019$  \\
			$\mathrm{BIC}$ & $1612.87377$ & $1628.03168$ & $1660.95996$ & $4.39464$  \\
			\hline\hline
	\end{tabular}
	\label{tab:results_flat_RLCDM}
\end{table}
\begin{figure}[H]
	\centering
	\includegraphics[width=0.8\linewidth]{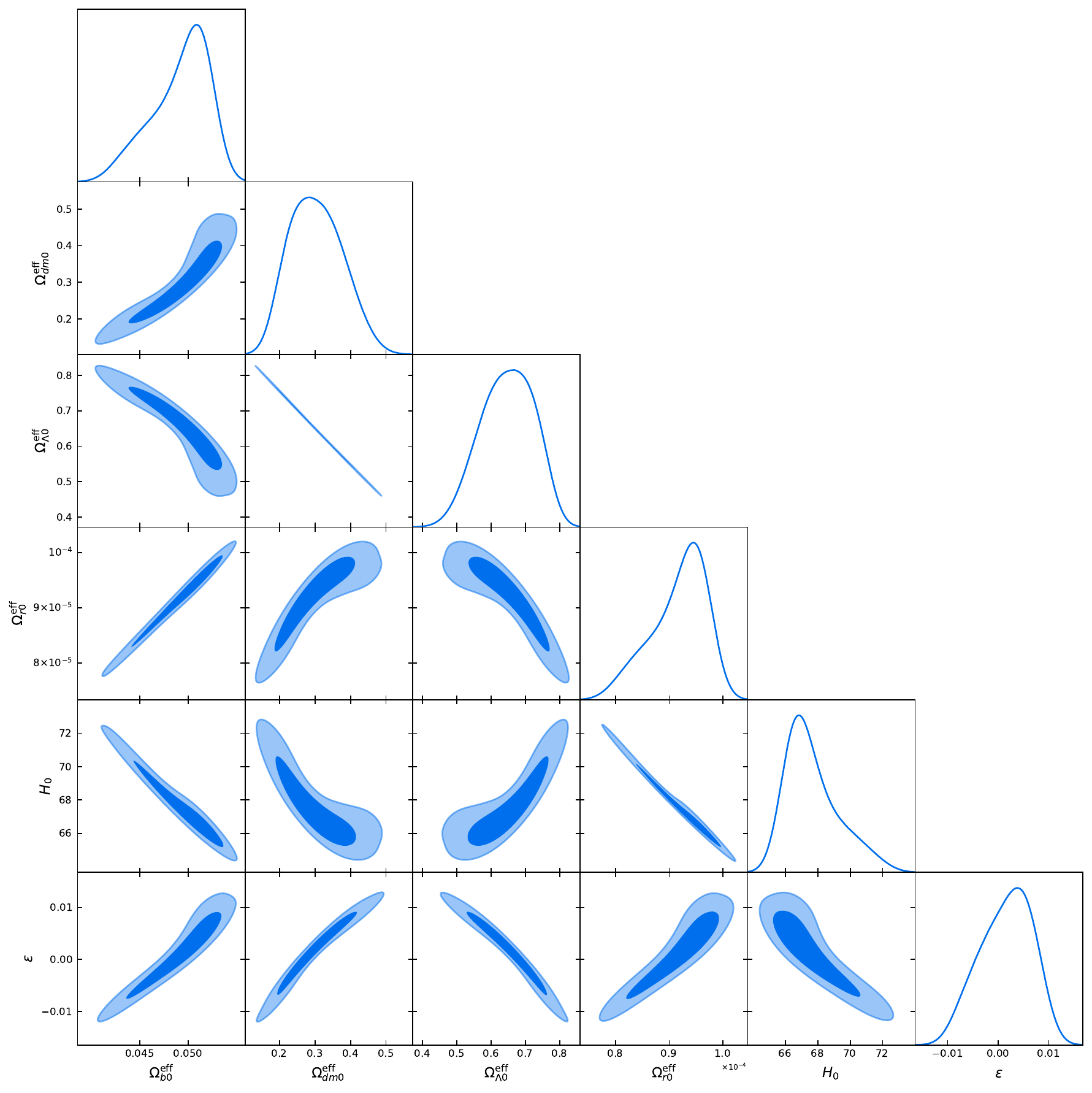}
	\caption{\small 
		Triangle plot for the flat R-$\Lambda$CDM model using the Planck 2018 CMB distance priors (D$_4$ dataset). The additional parameter $\epsilon$ introduces a clear degeneracy with $H_0$, leading to a broader allowed parameter space. This degeneracy illustrates how small deviations from the standard conservation law can affect the inferred value of the Hubble constant.
	}
	\label{fig:triangle_flat_RLCDM_D4}
\end{figure}
Figure~\ref{fig:triangle_flat_RLCDM_D4} presents the corresponding posterior distributions for the flat R-$\Lambda$CDM model. Compared to the standard case, the inclusion of the Rastall parameter $\epsilon$ leads to a noticeable broadening of the parameter space and introduces additional degeneracies, most prominently between $\epsilon$ and $H_0$.

The contour structure clearly indicates that small positive values of $\epsilon$ are associated with slightly higher values of the Hubble constant, providing a direct visual manifestation of the $\epsilon$--$H_0$ degeneracy discussed in this work. At the same time, the Planck-only dataset strongly constrains $\epsilon$ to remain close to zero, limiting the magnitude of the deviation from the standard $\Lambda$CDM scenario.

This behavior highlights that, although the Rastall framework allows for additional flexibility in the background dynamics, the high precision of CMB observations restricts the allowed deviations, resulting in only a mild shift in the inferred value of $H_0$.

\subsubsection*{Model Selection via AIC and BIC}

To assess the relative statistical performance of the flat R-$\Lambda$CDM model, we compare the AIC and BIC values obtained for dataset combinations D$_1$, D$_2$, and D$_3$ (Table~\ref{tab:AIC_BIC_comparison_flat}). The differences are defined as $\Delta$AIC $=\mathrm{AIC}_{\rm R}-\mathrm{AIC}_{\Lambda}$ and similarly for $\Delta$BIC.

For D$_1$, the R-$\Lambda$CDM model yields $\Delta$AIC$=-1.52$, indicating a marginal preference over the standard $\Lambda$CDM model according to the usual $|\Delta \mathrm{AIC}|<2$ criterion. However, the corresponding $\Delta$BIC$=+3.96$ suggests that the additional parameter $\epsilon$ is moderately penalized once model complexity is taken into account.

For D$_2$, both information criteria favor the standard $\Lambda$CDM model. The AIC difference is $\Delta$AIC$=+1.78$, indicating only weak statistical evidence against the R-$\Lambda$CDM model, while the BIC difference $\Delta$BIC$=+7.27$ indicates strong evidence against the more complex R-$\Lambda$CDM scenario within this dataset configuration.

A markedly different behavior is observed for D$_3$. In this case, the R-$\Lambda$CDM model is strongly favored over the standard scenario, with $\Delta$AIC$=-14.88$ and $\Delta$BIC$=-9.40$. According to conventional model selection thresholds, these values correspond to decisive statistical support for the R-$\Lambda$CDM model when BAO and CMB data are included simultaneously.

This preference can be interpreted as a direct consequence of the additional flexibility introduced by the Rastall parameter $\epsilon$, which allows the model to better accommodate the combined constraints from late- and early-Universe observables, particularly through its impact on the background expansion history.

It is important to emphasize that the absolute AIC and BIC values are meaningful only within a fixed dataset configuration. Therefore, comparisons should be interpreted separately for each dataset combination rather than across D$_1$, D$_2$, and D$_3$.

We also note that the D$_4$ (Planck-only) dataset is not included in this comparison, since information criteria are not directly comparable across datasets with substantially different effective numbers of data points, and such a comparison would therefore not be statistically meaningful.

From the perspective of the Hubble tension, these results further reinforce the dataset-dependent nature of the inferred cosmological parameters. In particular, the strong statistical preference for the R-$\Lambda$CDM model in D$_3$ correlates with the regime where small positive values of $\epsilon$ are favored and the inferred value of $H_0$ is shifted toward intermediate values between the Planck and SH0ES determinations. This behavior highlights that the apparent improvement in model performance is closely tied to the interplay between early- and late-Universe datasets, rather than representing a universal resolution of the tension.

Overall, the information criteria analysis indicates that the statistical viability of the R-$\Lambda$CDM model strongly depends on dataset composition. While it is moderately disfavored in D$_2$, it receives substantial support in D$_3$, highlighting the key role of early-Universe constraints in assessing deviations from the standard $\Lambda$CDM dynamics.

\begin{table}[h]
	\centering
	\caption{\small Comparison of AIC and BIC values between flat-universe R-$\Lambda$CDM and $\Lambda$CDM models for datasets D$_1$, D$_2$, and D$_3$. The differences are defined as $\Delta\mathrm{AIC}=\mathrm{AIC}_{\rm R}-\mathrm{AIC}_{\Lambda}$ and similarly for $\Delta\mathrm{BIC}$. Negative values indicate a preference for the R-$\Lambda$CDM model. The comparison is performed within each dataset independently, as AIC and BIC are not directly comparable across datasets with different numbers of data points.}
	\setlength{\tabcolsep}{6pt}
	\renewcommand{\arraystretch}{1.1}
		\begin{tabular}{l
				c c c @{\hskip 25pt}
				c c c}
			\hline\hline
			\textbf{Dataset} 
			& \multicolumn{3}{c}{\textbf{AIC}} 
			& \multicolumn{3}{c}{\textbf{BIC}} \\
			\cline{2-4} \cline{5-7}
			& $\Lambda$CDM & R-$\Lambda$CDM & $\Delta$AIC
			& $\Lambda$CDM & R-$\Lambda$CDM & $\Delta$BIC \\
			\hline
			D$_1$ & 1581.53 & 1580.01 & $-1.52$
			& 1608.92 & 1612.87 & $+3.96$ \\
			
			D$_2$ & 1593.34 & 1595.12 & $+1.78$
			& 1620.76 & 1628.03 & $+7.27$ \\
			
			D$_3$ & 1642.93 & 1628.04 & $-14.88$
			& 1670.36 & 1660.96 & $-9.40$ \\
			\hline\hline
	\end{tabular}
	\label{tab:AIC_BIC_comparison_flat}
\end{table}

\subsubsection{R-$\Lambda$CDM model in an open universe}

\begin{figure}[H]
	\centering
	\mbox{\includegraphics[width=0.8\textwidth]{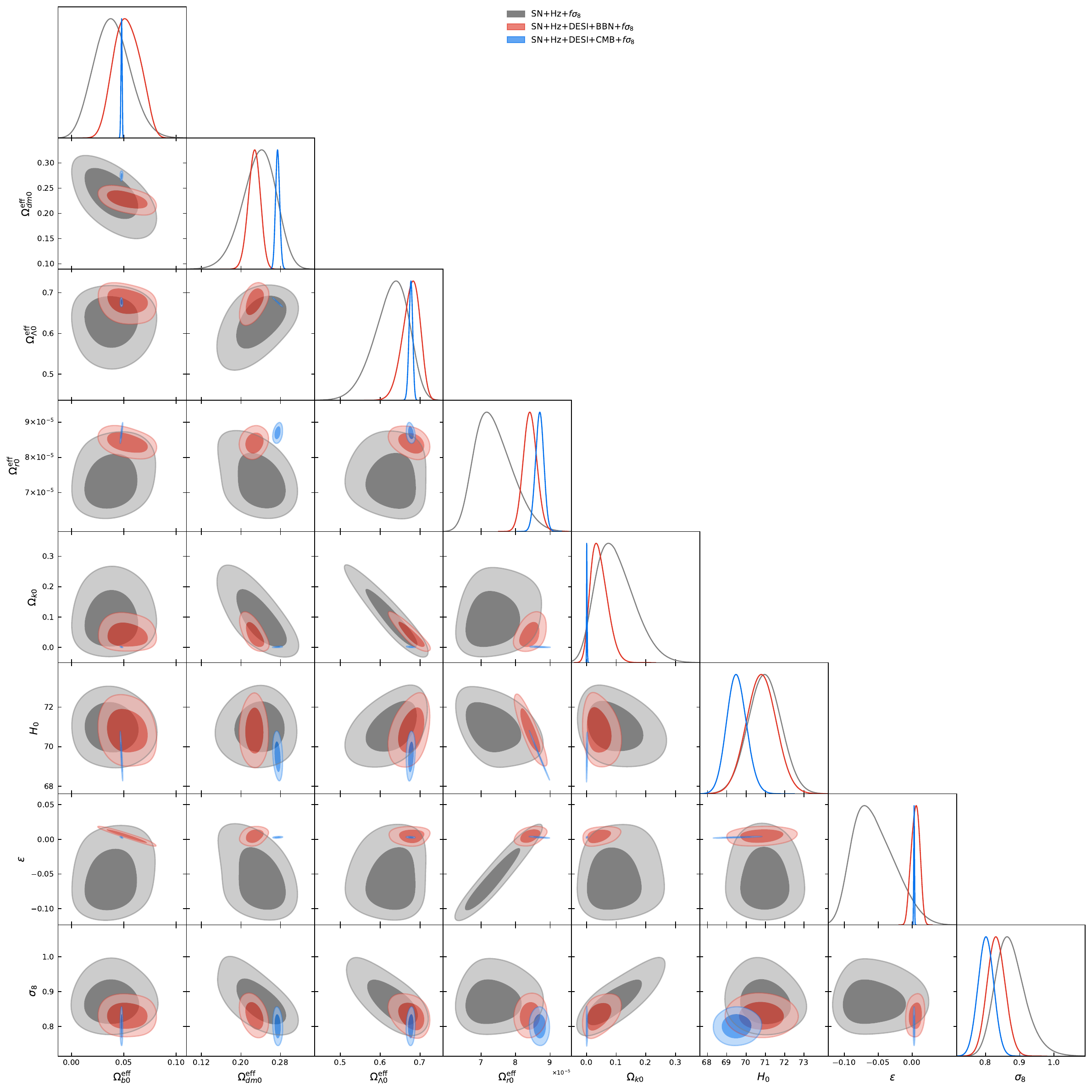}}
	\caption{\small One-dimensional marginalized likelihoods and $1\sigma$ and $2\sigma$ confidence contours of the open R-$\Lambda$CDM ($\Omega_{k_0}>0$) model parameters inferred from D$_1$ (SN+$H(z)$+$f\sigma_8$), D$_2$ (SN+$H(z)$+DESI+BBN+$f\sigma_8$), and D$_3$ (SN+$H(z)$+DESI+CMB priors+$f\sigma_8$).}
	\label{fig:rlcdm_nonflat_fig1_k>0}
\end{figure}

The cosmological parameter estimates for the open R-$\Lambda$CDM model, obtained using datasets D$_1$, D$_2$, and D$_3$, are summarized in Table~\ref{tab:results_open_RLCDM} and illustrated in Figure~\ref{fig:rlcdm_nonflat_fig1_k>0}. In this framework, effective density parameters $\Omega_i^{\rm eff}$ govern the background expansion, while the curvature parameter $\Omega_{k0}$ remains unrescaled.

The Hubble constant $H_0$ is estimated to be $70.96 \pm 0.84$, $70.79 \pm 0.76$, and $69.50 \pm 0.51$ km s$^{-1}$ Mpc$^{-1}$ for datasets D$_1$, D$_2$, and D$_3$, respectively. These values lie between the Planck 2018 reference ($67.4 \pm 0.5$ km s$^{-1}$ Mpc$^{-1}$) and the local SH0ES measurement ($73.2 \pm 1.3$ km s$^{-1}$ Mpc$^{-1}$).

In the standard flat $\Lambda$CDM model, the tension between Planck and SH0ES is approximately $4.17\sigma$. In the open R-$\Lambda$CDM model, the tension with Planck is reduced to about $3.63\sigma$, $3.73\sigma$, and $2.92\sigma$ for D$_1$, D$_2$, and D$_3$, respectively. Similarly, the tension with SH0ES is approximately $1.45\sigma$, $1.61\sigma$, and $2.64\sigma$. Thus, the open R-$\Lambda$CDM framework shifts $H_0$ toward intermediate values, with D$_3$ moving toward the early-Universe calibration when CMB information is included.

Regarding other parameters, the effective dark matter density $\Omega_{dm0}^{\rm eff}$ is around $0.24$ for D$_1$ and D$_2$, while D$_3$ favors a higher value near $0.274$. The effective baryonic density $\Omega_{b0}^{\rm eff}$ remains broadly consistent across datasets, with the tightest constraint obtained for D$_3$.

The spatial curvature parameter $\Omega_{k0}$ is positive in all cases but exhibits substantial dataset dependence. D$_1$ allows the largest curvature, $\Omega_{k0} \sim 0.09$, D$_2$ prefers a smaller value around $0.04$, and D$_3$ is consistent with a nearly flat Universe, with $\Omega_{k0}$ close to zero and tightly constrained.

The effective dark energy density $\Omega_{\Lambda0}^{\rm eff}$ lies in the range $0.63$–$0.68$, with D$_1$ yielding the lowest value and D$_2$ and D$_3$ showing mutually consistent results. The effective radiation density $\Omega_{r0}^{\rm eff}$ remains at the expected order of $10^{-5}$ in all cases, with tighter constraints obtained when BAO and CMB information are included.

The Rastall deviation parameter $\epsilon$ exhibits a distinct pattern. Dataset D$_1$ prefers a negative value, $\epsilon \approx -0.061$, corresponding to roughly a $2\sigma$ deviation from zero, whereas D$_2$ and D$_3$ yield values consistent with $\epsilon=0$ within $1\sigma$. This indicates that the inclusion of BAO and especially CMB data drives the model closer to the standard $\Lambda$CDM limit.

The structure growth parameter $\sigma_8$ decreases from approximately $0.87$ in D$_1$ to $0.80$ in D$_3$, with D$_2$ taking an intermediate value.

Finally, the information criteria show that D$_1$ yields the lowest AIC and BIC values within the open R-$\Lambda$CDM scenario, while D$_3$ produces the largest values. Since the dataset compositions differ, these criteria should be interpreted separately within each configuration.

\subsubsection*{Comparison of dataset D$_1$ with D$_2$ and D$_3$}

Comparing D$_1$ and D$_2$ reveals moderate deviations in $\Omega_{k0}$ ($0.7\sigma$), $\epsilon$ ($2.0\sigma$), and $\Omega_{r0}^{\rm eff}$ ($2.0\sigma$), while $\Omega_{dm0}^{\rm eff}$, $\Omega_{\Lambda0}^{\rm eff}$, $H_0$, and $\sigma_8$ remain statistically consistent within $1\sigma$. Overall, most parameters are compatible within uncertainties.

In contrast, comparing D$_1$ and D$_3$ shows stronger differences in curvature and dark matter density. The curvature parameter $\Omega_{k0}$ differs at approximately the $1.4\sigma$ level, $\Omega_{dm0}^{\rm eff}$ shifts by about $1.1\sigma$, and $\sigma_8$ differs by roughly $1.6\sigma$. The Hubble constant differs by approximately $1.5\sigma$. Despite these variations, the parameter sets remain statistically compatible within the open R-$\Lambda$CDM framework.

\begin{table}[h]
	\caption{\small The 68\% confidence limits of open R-$\Lambda$CDM cosmological parameters from D$_1$, D$_2$, and D$_3$ datasets. $H_0$ is in km\,s$^{-1}$\,Mpc$^{-1}$. In Rastall gravity we report effective density parameters $\Omega_i^{\rm eff}\equiv\tilde{\Omega}_i$ that govern the background expansion; $\Omega_{k0}$ is not rescaled.}
	\centering
	\setlength{\tabcolsep}{9pt}
	\renewcommand{\arraystretch}{1.4}
		\begin{tabular}{lccc}
			\hline\hline
			\textbf{CPs} & \textbf{D$_1$} & \textbf{D$_2$} & \textbf{D$_3$} \\
			\hline
			$\Omega_{b0}^{\rm eff}$ & $0.03805^{+0.01746}_{-0.01656}$ & $0.05243^{+0.01400}_{-0.01240}$ & $0.04772^{+0.00050}_{-0.00056}$ \\
			$\Omega_{dm0}^{\rm eff}$ & $0.23881^{+0.03134}_{-0.03577}$ & $0.22733^{+0.01153}_{-0.01236}$ & $0.27424^{+0.00413}_{-0.00398}$ \\
			$\Omega_{\Lambda0}^{\rm eff}$ & $0.63181^{+0.03810}_{-0.04785}$ & $0.67908^{+0.01962}_{-0.02389}$ & $0.67675^{+0.00438}_{-0.00469}$ \\
			$\Omega_{r0}^{\rm eff}$ & $(7.29^{+0.63}_{-0.47}) \times 10^{-5}$ & $(8.42^{+0.20}_{-0.19}) \times 10^{-5}$ & $(8.70 \pm 0.12) \times 10^{-5}$ \\
			$\Omega_{k0}$ & $0.08911^{+0.07347}_{-0.05727}$ & $0.03931^{+0.03211}_{-0.02515}$ & $0.00108^{+0.00120}_{-0.00076}$ \\
			$H_0$ & $70.96 \pm 0.84$ & $70.79 \pm 0.76$ & $69.50^{+0.52}_{-0.49}$ \\
			$\epsilon$ & $-0.06077^{+0.03640}_{-0.02718}$ & $0.00526^{+0.00604}_{-0.00697}$ & $0.00296^{+0.00053}_{-0.00054}$ \\
			$\sigma_8$ & $0.8685^{+0.0473}_{-0.0374}$ & $0.8307^{+0.0263}_{-0.0254}$ & $0.8009^{+0.0224}_{-0.0226}$ \\
			\hline
			$\mathrm{AIC}$ & $1581.62349$ & $1596.38812$ & $1621.95742$ \\
			$\mathrm{BIC}$ & $1619.96276$ & $1634.78263$ & $1660.35979$ \\
			\hline\hline
		\end{tabular}
	\label{tab:results_open_RLCDM}
\end{table}
\subsubsection*{Model Selection via AIC and BIC}

The relative statistical performance of the open R-$\Lambda$CDM model was evaluated using the Akaike Information Criterion (AIC) and the Bayesian Information Criterion (BIC) for dataset combinations D$_1$, D$_2$, and D$_3$ (Table~\ref{tab:AIC_BIC_comparison_open}). The differences are defined as $\Delta$AIC$=\mathrm{AIC}_{\rm R}-\mathrm{AIC}_{\Lambda}$ and similarly for $\Delta$BIC.

For D$_1$, the R-$\Lambda$CDM model yields $\Delta$AIC$=-0.59$, indicating statistical equivalence with the standard $\Lambda$CDM model according to the conventional $|\Delta \mathrm{AIC}|<2$ criterion. However, the corresponding $\Delta$BIC$=+4.89$ shows a moderate penalty for the additional parameter when model complexity is more strongly weighted.

For D$_2$, the AIC difference ($\Delta$AIC$=+1.38$) indicates statistical equivalence between the two models, while the BIC difference ($\Delta$BIC$=+6.87$) provides strong evidence in favor of the standard $\Lambda$CDM model due to the stronger penalty on model complexity.

A qualitatively different behavior is observed for D$_3$. In this case, the open R-$\Lambda$CDM model significantly outperforms the standard $\Lambda$CDM model, with $\Delta$AIC$=-25.60$ and $\Delta$BIC$=-20.11$. According to conventional model selection thresholds, such large negative differences correspond to strong statistical evidence in favor of the R-$\Lambda$CDM scenario when BAO and CMB data are simultaneously included.

As emphasized above, the absolute values of AIC and BIC are meaningful only within a fixed dataset configuration.

Overall, the information criteria analysis indicates that the statistical viability of the open R-$\Lambda$CDM model depends sensitively on dataset composition. While it remains statistically equivalent to $\Lambda$CDM in D$_1$ and statistically equivalent under AIC but disfavored by BIC in D$_2$, it receives decisive support in D$_3$, highlighting the crucial role of CMB constraints in assessing deviations from the standard cosmological model.

\begin{table}[h]
	\centering
	\caption{\small Comparison of AIC and BIC values between open-universe R-$\Lambda$CDM and $\Lambda$CDM models for datasets D$_1$, D$_2$, and D$_3$. The differences are defined as $\Delta$AIC$=\mathrm{AIC}_{\rm R}-\mathrm{AIC}_{\Lambda}$ and similarly for $\Delta$BIC.}
	\setlength{\tabcolsep}{6pt}
	\renewcommand{\arraystretch}{1.1}
		\begin{tabular}{l
				c c c @{\hskip 25pt}
				c c c}
			\hline\hline
			\textbf{Dataset} 
			& \multicolumn{3}{c}{\textbf{AIC}} 
			& \multicolumn{3}{c}{\textbf{BIC}} \\
			\cline{2-4} \cline{5-7}
			& $\Lambda$CDM & R-$\Lambda$CDM & $\Delta$AIC
			& $\Lambda$CDM & R-$\Lambda$CDM & $\Delta$BIC \\
			\hline
			D$_1$ & 1582.21 & 1581.62 & $-0.59$
			& 1615.07 & 1619.96 & $+4.89$ \\
			
			D$_2$ & 1595.00 & 1596.39 & $+1.38$
			& 1627.91 & 1634.78 & $+6.87$ \\
			
			D$_3$ & 1647.56 & 1621.96 & $-25.60$
			& 1680.47 & 1660.36 & $-20.11$ \\
			\hline\hline
	\end{tabular}
	\label{tab:AIC_BIC_comparison_open}
\end{table}

\subsubsection{R-$\Lambda$CDM model in a closed universe}

\begin{figure}[H]
	\centering
	\mbox{\includegraphics[width=0.8\textwidth]{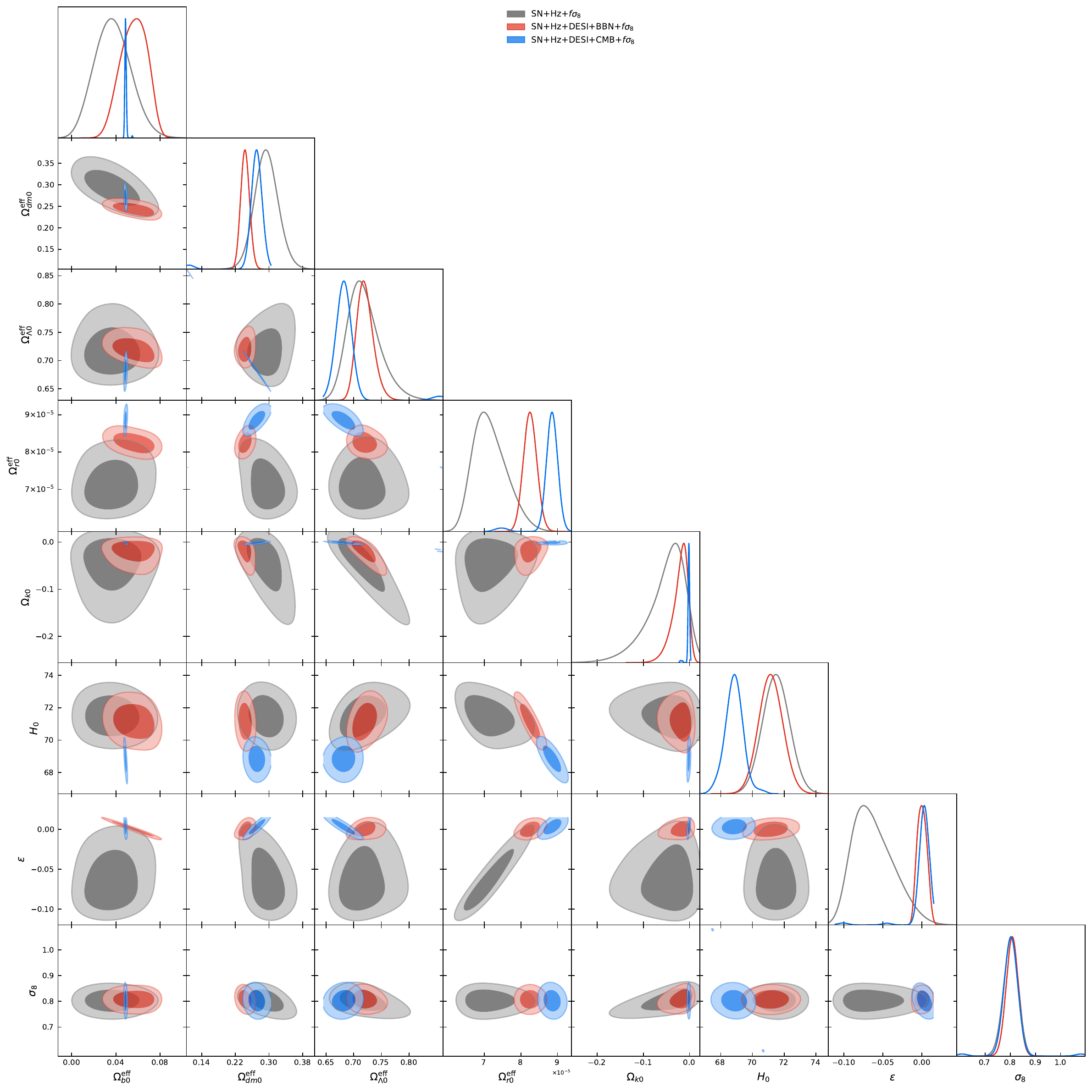}}
	\caption{\small One-dimensional marginalized likelihoods and $1\sigma$ and $2\sigma$ confidence contours of the closed R-$\Lambda$CDM ($\Omega_{k_0}<0$) model parameters inferred from D$_1$ (SN+$H(z)$+$f\sigma_8$), D$_2$ (SN+$H(z)$+DESI+BBN+$f\sigma_8$), and D$_3$ (SN+$H(z)$+DESI+CMB priors+$f\sigma_8$).}
	\label{fig:rlcdm_nonflat_fig1_k<0}
\end{figure}

Table~\ref{tab:results_closed_RLCDM} summarizes the cosmological parameter estimates of the closed R-$\Lambda$CDM model obtained from the three independent observational datasets D$_1$, D$_2$, and D$_3$. The corresponding likelihood contours are shown in Figure~\ref{fig:rlcdm_nonflat_fig1_k<0}. In this framework, the effective density parameters $\Omega_i^{\rm eff}$ govern the background expansion, while the curvature parameter $\Omega_{k0}$ remains unrescaled.

The Hubble constant is constrained to be $H_0 = 71.48 \pm 0.84$, $71.15 \pm 0.75$, and $68.86 \pm 0.54$ km s$^{-1}$ Mpc$^{-1}$ for D$_1$, D$_2$, and D$_3$, respectively (symmetrized $1\sigma$ uncertainties). Relative to the Planck 2018 determination ($67.4 \pm 0.5$ km s$^{-1}$ Mpc$^{-1}$), the corresponding tensions are approximately $4.19\sigma$, $4.16\sigma$, and $1.99\sigma$. Compared to the local SH0ES measurement ($73.2 \pm 1.3$ km s$^{-1}$ Mpc$^{-1}$), the tensions are $1.11\sigma$, $1.37\sigma$, and $3.09\sigma$, respectively. Thus, in the absence of CMB information (D$_1$ and D$_2$), the closed R-$\Lambda$CDM model favors values of $H_0$ closer to the local distance-ladder measurement, whereas the inclusion of CMB data in D$_3$ shifts the result toward the Planck value and reduces the Planck tension to the $\sim 2\sigma$ level.

The effective dark matter density parameter $\Omega_{dm0}^{\rm eff}$ ranges from $0.243$ (D$_2$) to $0.293$ (D$_1$), with D$_3$ yielding an intermediate value of $0.270$. The effective baryon density $\Omega_{b0}^{\rm eff}$ lies between $0.036$ and $0.057$, with the tightest constraint obtained for D$_3$.

The spatial curvature parameter $\Omega_{k0}$ is negative in all cases, confirming a closed spatial geometry. Its magnitude decreases as additional early-Universe information is incorporated: $\Omega_{k0} \approx -0.039$ (D$_1$), $-0.016$ (D$_2$), and $-0.0013$ (D$_3$). The D$_3$ result indicates only a mild ($\sim1\sigma$) preference for negative curvature and remains close to spatial flatness, highlighting the strong constraining power of CMB data on geometry.

The effective dark energy density $\Omega_{\Lambda0}^{\rm eff}$ varies between $0.68$ and $0.72$, while the effective radiation density $\Omega_{r0}^{\rm eff}$ remains of order $10^{-5}$ in all cases, with progressively tighter uncertainties from D$_1$ to D$_3$.

The Rastall deviation parameter $\epsilon$ shows a clear dataset dependence. D$_1$ favors a negative value, $\epsilon = -0.067 \pm 0.027$, corresponding to roughly a $2\sigma$ deviation from zero. In contrast, D$_2$ yields a value fully consistent with $\epsilon = 0$ within $1\sigma$, whereas D$_3$ prefers a small but statistically significant positive value, $\epsilon \simeq 0.0023$ ($\sim 3.8\sigma$). This indicates that the inclusion of BAO and especially CMB information drives the model closer to (or tightly constrains deviations from) the standard $\Lambda$CDM limit.

The structure growth parameter $\sigma_8$ remains stable across datasets, with values around $0.80$ and differences well within $1\sigma$.

The information criteria indicate that D$_1$ provides the lowest AIC and BIC values within the closed R-$\Lambda$CDM scenario, followed by D$_2$, while D$_3$ yields the largest values. Since the dataset compositions differ, these criteria should be interpreted separately within each configuration.
\subsubsection*{Comparison of dataset D$_1$ with D$_2$ and D$_3$}

The comparison between D$_1$ and D$_2$ reveals a significant shift in the effective dark matter density $\Omega_{dm0}^{\rm eff}$ at the $1.73\sigma$ level. The effective baryon density differs by approximately $0.98\sigma$, while the curvature parameter $\Omega_{k0}$ shifts by about $0.54\sigma$. The Hubble constant remains statistically consistent ($0.29\sigma$), and $\sigma_8$ differs by only $0.13\sigma$. The radiation density parameter shows a deviation of approximately $2.33\sigma$, and the Rastall parameter $\epsilon$ differs by about $2.40\sigma$. Overall, most parameters remain compatible within $2\sigma$.

A comparison between D$_1$ and D$_3$ indicates moderate deviations in several parameters. The dark matter density differs by approximately $0.85\sigma$, the baryon density by about $0.75\sigma$, and the curvature parameter by roughly $0.96\sigma$. The Hubble constant shifts by approximately $2.64\sigma$, while $\sigma_8$ remains essentially unchanged ($0.01\sigma$). The radiation density parameter shows a deviation close to $3.65\sigma$, reflecting its sensitivity to early-Universe constraints. The Rastall parameter differs by about $2.56\sigma$, driven mainly by the negative value preferred in D$_1$.

These results indicate that, although several parameters exhibit moderate shifts when CMB information is included, the overall parameter sets remain statistically compatible within the closed R-$\Lambda$CDM framework.

\begin{table}[h]
	\caption{\small The 68\% confidence limits of closed R-$\Lambda$CDM cosmological parameters from D$_1$, D$_2$, and D$_3$ datasets. $H_0$ is in km\,s$^{-1}$\,Mpc$^{-1}$. In Rastall gravity we report effective density parameters $\Omega_i^{\rm eff}\equiv\tilde{\Omega}_i$ that govern the background expansion; $\Omega_{k0}$ is not rescaled.}
	\centering
	\setlength{\tabcolsep}{9pt}
	\renewcommand{\arraystretch}{1.5}
		\begin{tabular}{lccc}
			\hline\hline
			\textbf{CPs} & \textbf{D$_1$} & \textbf{D$_2$} & \textbf{D$_3$} \\
			\hline
			$\Omega_{b0}^{\rm eff}$ & $0.03646^{+0.01664}_{-0.01589}$ & $0.05684^{+0.01231}_{-0.01328}$ & $0.04870^{+0.00068}_{-0.00054}$ \\
			$\Omega_{dm0}^{\rm eff}$ & $0.29318^{+0.02764}_{-0.02649}$ & $0.24318^{+0.01044}_{-0.01010}$ & $0.26989^{+0.00407}_{-0.00437}$ \\
			$\Omega_{\Lambda0}^{\rm eff}$ & $0.71445^{+0.03240}_{-0.02386}$ & $0.71924^{+0.01589}_{-0.01324}$ & $0.68274^{+0.00498}_{-0.00452}$ \\
			$\Omega_{r0}^{\rm eff}$ & $(7.09^{+0.54}_{-0.39}) \times 10^{-5}$ & $(8.25 \pm 0.18) \times 10^{-5}$ & $(8.85 \pm 0.13) \times 10^{-5}$ \\
			$\Omega_{k0}$ & $-0.03897^{+0.02832}_{-0.04994}$ & $-0.01614^{+0.01182}_{-0.02057}$ & $-0.00129^{+0.00091}_{-0.00148}$ \\
			$H_0$ & $71.48^{+0.84}_{-0.83}$ & $71.15^{+0.76}_{-0.74}$ & $68.86^{+0.53}_{-0.54}$ \\
			$\epsilon$ & $-0.06681^{+0.03127}_{-0.02279}$ & $-0.00009^{+0.00660}_{-0.00645}$ & $0.00230^{+0.00057}_{-0.00064}$ \\
			$\sigma_8$ & $0.8030^{+0.0276}_{-0.0290}$ & $0.8078^{+0.0231}_{-0.0237}$ & $0.8025^{+0.0231}_{-0.0230}$ \\
			\hline
			$\mathrm{AIC}$ & $1582.06061$ & $1597.27448$ & $1621.92194$ \\
			$\mathrm{BIC}$ & $1620.39988$ & $1635.66899$ & $1660.32431$ \\
			\hline\hline
	\end{tabular}
	\label{tab:results_closed_RLCDM}
\end{table}
\subsubsection*{Model Selection via AIC and BIC}

Table~\ref{tab:AIC_BIC_comparison_closed} presents a comparative assessment of the Akaike Information Criterion (AIC) and the Bayesian Information Criterion (BIC) for the closed R-$\Lambda$CDM and standard closed $\Lambda$CDM models. The differences are defined as $\Delta$AIC$=\mathrm{AIC}_{\rm R}-\mathrm{AIC}_{\Lambda}$ and similarly for $\Delta$BIC.

For dataset D$_1$, the R-$\Lambda$CDM model yields $\Delta$AIC$=-1.65$, indicating statistical equivalence with the standard $\Lambda$CDM model according to the conventional $|\Delta \mathrm{AIC}|<2$ criterion. However, the corresponding $\Delta$BIC$=+3.82$ reflects the penalty imposed by the additional Rastall parameter when model complexity is more strongly weighted.

For D$_2$, the AIC difference ($\Delta$AIC$=+1.82$) indicates statistical equivalence between the two models, while the BIC difference ($\Delta$BIC$=+7.31$) provides strong evidence in favor of the standard $\Lambda$CDM model.

A markedly different behavior is observed for D$_3$. In this case, the closed R-$\Lambda$CDM model significantly outperforms the standard $\Lambda$CDM model, with $\Delta$AIC$=-12.59$ and $\Delta$BIC$=-7.10$. According to standard model selection thresholds, such large negative differences correspond to strong statistical evidence in favor of the R-$\Lambda$CDM extension when BAO and CMB data are simultaneously included.

As emphasized above, the absolute values of AIC and BIC are meaningful only within a fixed dataset configuration.

Overall, the information criteria analysis indicates that the statistical viability of the closed R-$\Lambda$CDM model depends sensitively on dataset composition. While it remains statistically equivalent to $\Lambda$CDM in D$_1$ and statistically equivalent under AIC but disfavored by BIC in D$_2$, it receives strong support in D$_3$, highlighting the crucial role of CMB constraints in assessing deviations from the standard cosmological model.

\begin{table}[h]
	\centering
	\caption{\small Comparison of AIC and BIC values between closed-universe R-$\Lambda$CDM and $\Lambda$CDM models for datasets D$_1$, D$_2$, and D$_3$. The differences are defined as $\Delta$AIC$=\mathrm{AIC}_{\rm R}-\mathrm{AIC}_{\Lambda}$ and similarly for $\Delta$BIC.}
	\setlength{\tabcolsep}{6pt}
	\renewcommand{\arraystretch}{1.1}
		\begin{tabular}{l
				c c c @{\hskip 25pt}
				c c c}
			\hline\hline
			\textbf{Dataset} 
			& \multicolumn{3}{c}{\textbf{AIC}} 
			& \multicolumn{3}{c}{\textbf{BIC}} \\
			\cline{2-4} \cline{5-7}
			& $\Lambda$CDM & R-$\Lambda$CDM & $\Delta$AIC
			& $\Lambda$CDM & R-$\Lambda$CDM & $\Delta$BIC \\
			\hline
			D$_1$ & 1583.72 & 1582.06 & $-1.65$
			& 1616.58 & 1620.40 & $+3.82$ \\
			
			D$_2$ & 1595.45 & 1597.27 & $+1.82$
			& 1628.36 & 1635.67 & $+7.31$ \\
			
			D$_3$ & 1634.51 & 1621.92 & $-12.59$
			& 1667.42 & 1660.32 & $-7.10$ \\
			\hline\hline
	\end{tabular}
	\label{tab:AIC_BIC_comparison_closed}
\end{table}

\subsubsection{Assessment of geometrical models based on parameter consistency: The R-$\Lambda$CDM case}

Table~\ref{tabTre} quantifies the internal consistency of the R-$\Lambda$CDM model across different dataset combinations and spatial geometries. The cosmological parameter differences (CPDs), expressed in units of the combined statistical uncertainty, provide a direct measure of the stability of the inferred parameters when progressively incorporating additional observational information.

A common feature across all geometries is that the largest dataset-induced shifts occur in the effective radiation density, $\Omega_{r0}^{\rm eff}$, and in the Rastall deviation parameter, $\epsilon$. In the flat case, $\Omega_{r0}^{\rm eff}$ exhibits a substantial shift of $4.03\sigma$ between D$_1$ and D$_3$, while $\epsilon$ varies at the $\sim 2.3\sigma$ level. A similarly pronounced CMB-driven effect is observed in the Hubble constant, which changes by $3.07\sigma$ between D$_1$ and D$_3$. It is important to note that the shift in $\Omega_{r0}^{\rm eff}$ primarily reflects its implicit dependence on $H_0$, rather than a modification of radiation physics itself. These results indicate that, in the flat R-$\Lambda$CDM scenario, the inclusion of early-Universe information significantly affects both the radiation sector and the inferred expansion rate.

In contrast, the open R-$\Lambda$CDM geometry exhibits the most homogeneous internal behavior. No parameter exceeds the $3\sigma$ level for any dataset comparison. The largest deviations are found in $\Omega_{r0}^{\rm eff}$ ($2.50\sigma$ for $|D_1-D_3|$) and in $\epsilon$ (approximately $2.0\sigma$), while the Hubble constant remains comparatively stable, with a shift of $1.49\sigma$ between D$_1$ and D$_3$. The curvature parameter $\Omega_{k0}$ itself varies at the $\lesssim 1.35\sigma$ level. Overall, the open geometry provides the most internally uniform parameter reconstruction within the R-$\Lambda$CDM framework.

The closed R-$\Lambda$CDM case exhibits intermediate behavior. While the growth amplitude $\sigma_8$ is remarkably stable ($0.01\sigma$ between D$_1$ and D$_3$), the effective radiation density still shows a significant $3.65\sigma$ shift, and the Hubble constant varies by $2.64\sigma$ when CMB information is included. The Rastall parameter also displays a persistent $\sim 2.5\sigma$ dataset dependence. Therefore, although the closed geometry maintains strong consistency in the late-time growth sector, it remains sensitive to early-Universe constraints in the radiation and expansion-rate sectors.

Taken together, the CPD analysis indicates that the dominant source of dataset sensitivity in R-$\Lambda$CDM is associated with $\Omega_{r0}^{\rm eff}$ and, through its coupling, the inferred value of $H_0$. Among the three geometrical scenarios, the open R-$\Lambda$CDM model exhibits the most homogeneous internal parameter behavior, as it avoids deviations exceeding the $3\sigma$ level while maintaining stable estimates across matter, curvature, and expansion parameters.

\begin{table}[h]
	\caption{\small Updated cosmological parameter differences (CPD) expressed in units of $\sigma$ for the R-$\Lambda$CDM model in flat and non-flat geometries. The CPDs are computed using symmetrized $1\sigma$ uncertainties, $\sigma=(\sigma_{+}+\sigma_{-})/2$, and $\sigma_{\rm comb}=\sqrt{\sigma_1^2+\sigma_2^2}$. In Rastall gravity, all density parameters are the effective ones $\Omega_i^{\rm eff}$ governing the background expansion, while $\Omega_{k0}$ is not rescaled.}
	\setlength{\tabcolsep}{4pt}
	\renewcommand{\arraystretch}{1.2}
	\begin{center}
			\begin{tabular}{l
					cc @{\hskip 10pt}
					cc @{\hskip 10pt}
					cc}
				\hline\hline
				\textbf{CPD}
				& \multicolumn{2}{c}{\textbf{Flat}}
				& \multicolumn{2}{c}{\textbf{Open}}
				& \multicolumn{2}{c}{\textbf{Closed}} \\
				& $|\textbf{D$_1$}-\textbf{D$_2$}|$
				& $|\textbf{D$_1$}-\textbf{D$_3$}|$
				& $|\textbf{D$_1$}-\textbf{D$_2$}|$
				& $|\textbf{D$_1$}-\textbf{D$_3$}|$
				& $|\textbf{D$_1$}-\textbf{D$_2$}|$
				& $|\textbf{D$_1$}-\textbf{D$_3$}|$ \\
				\hline
				$\Omega_{dm0}^{\rm eff}$      & $1.53\sigma$ & $<0.01\sigma$ & $0.32\sigma$ & $1.05\sigma$ & $1.73\sigma$ & $0.85\sigma$ \\
				$\Omega_{b0}^{\rm eff}$       & $0.85\sigma$ & $0.70\sigma$ & $0.67\sigma$ & $0.57\sigma$ & $0.98\sigma$ & $0.75\sigma$ \\
				$\Omega_{k0}$                 & --           & --           & $0.70\sigma$ & $1.35\sigma$ & $0.54\sigma$ & $0.96\sigma$ \\
				$\Omega_{\Lambda0}^{\rm eff}$ & $1.05\sigma$ & $0.73\sigma$ & $0.98\sigma$ & $1.04\sigma$ & $0.15\sigma$ & $1.11\sigma$ \\
				$\Omega_{r0}^{\rm eff}$       & $2.45\sigma$ & $4.03\sigma$ & $1.94\sigma$ & $2.50\sigma$ & $2.33\sigma$ & $3.65\sigma$ \\
				$\epsilon$                    & $2.26\sigma$ & $2.32\sigma$ & $2.03\sigma$ & $2.00\sigma$ & $2.40\sigma$ & $2.56\sigma$ \\
				$H_0$                         & $0.23\sigma$ & $3.07\sigma$ & $0.15\sigma$ & $1.49\sigma$ & $0.29\sigma$ & $2.64\sigma$ \\
				$\sigma_8$                    & $0.26\sigma$ & $0.70\sigma$ & $0.76\sigma$ & $1.41\sigma$ & $0.13\sigma$ & $0.01\sigma$ \\
				\hline\hline
		\end{tabular}
	\end{center}
	\label{tabTre}
\end{table}
\subsection{Dependence of $H_0$ constraints on dataset combinations}

A key aspect in the interpretation of the Hubble tension is the dependence of the inferred value of $H_0$ on the choice of observational datasets. In order to systematically investigate this issue, we consider four different dataset combinations, namely D$_1$, D$_2$, D$_3$, and D$_4$, where D$_4$ corresponds to the Planck 2018 CMB distance priors alone.

The results for all cosmological models are summarized in Table~\ref{tab:H0_summary_all_models}. A clear and robust trend emerges across all geometries and models. Late-time datasets (D$_1$ and D$_2$), which include supernovae, cosmic chronometers, and large-scale structure data, systematically favor higher values of the Hubble constant, typically in the range $H_0 \sim 70$--$71\ {\rm km\,s^{-1}\,Mpc^{-1}}$. In contrast, when CMB information is included (D$_3$), or when CMB data are used alone (D$_4$), the inferred value of $H_0$ shifts toward the Planck-preferred region, $H_0 \sim 67$--$68\ {\rm km\,s^{-1}\,Mpc^{-1}}$.

\begin{table}[t]
	\caption{\small Summary of the inferred Hubble constant $H_0$ (km\,s$^{-1}$\,Mpc$^{-1}$) for all investigated cosmological models and dataset combinations. Late-time datasets D$_1$ and D$_2$ generally favor higher values of $H_0$, while the inclusion of CMB information in D$_3$ and D$_4$ shifts the constraints toward the Planck-preferred region.}
	\centering
	\setlength{\tabcolsep}{9pt}
	\renewcommand{\arraystretch}{1.5}
		\begin{tabular}{lcccc}
			\hline\hline
			\textbf{Model} & \textbf{D$_1$} & \textbf{D$_2$} & \textbf{D$_3$} & \textbf{D$_4$} \\
			\hline
			
			Flat $\Lambda$CDM
			& $71.30^{+0.81}_{-0.79}$
			& $71.04^{+0.72}_{-0.73}$
			& $67.35 \pm 0.28$
			& $67.66^{+0.60}_{-0.61}$ \\
			
			Open $\Lambda$CDM
			& $70.92^{+0.84}_{-0.83}$
			& $70.75^{+0.75}_{-0.75}$
			& $67.55^{+0.29}_{-0.28}$
			& -- \\
			
			Closed $\Lambda$CDM
			& $71.43^{+0.84}_{-0.83}$
			& $71.18 \pm 0.74$
			& $67.18 \pm 0.30$
			& -- \\
			
			Flat R-$\Lambda$CDM
			& $71.32^{+0.84}_{-0.81}$
			& $71.06 \pm 0.75$
			& $68.50^{+0.40}_{-0.41}$
			& $67.13^{+2.28}_{-1.04}$ \\
			
			Open R-$\Lambda$CDM
			& $70.96 \pm 0.84$
			& $70.79 \pm 0.76$
			& $69.50^{+0.52}_{-0.49}$
			& -- \\
			
			Closed R-$\Lambda$CDM
			& $71.48^{+0.84}_{-0.83}$
			& $71.15^{+0.76}_{-0.74}$
			& $68.86^{+0.53}_{-0.54}$
			& -- \\
			
			\hline\hline
	\end{tabular}
	\label{tab:H0_summary_all_models}
\end{table}

This behavior is consistently observed in both $\Lambda$CDM and R--$\Lambda$CDM cosmologies, as well as for flat, open, and closed geometries. In particular, the D$_4$ dataset provides a direct estimate of $H_0$ from early-Universe information alone. For the flat $\Lambda$CDM model, we obtain $H_0 = 67.66^{+0.60}_{-0.61}\ {\rm km\,s^{-1}\,Mpc^{-1}}$, in full agreement with the Planck 2018 baseline result. Similarly, in the flat R--$\Lambda$CDM scenario we find $H_0 = 67.13^{+2.28}_{-1.04}\ {\rm km\,s^{-1}\,Mpc^{-1}}$, indicating that the inclusion of the Rastall parameter enlarges the allowed parameter space but does not significantly shift the central value when only CMB information is considered.

These results highlight that the Hubble tension can be interpreted as a discrepancy between early- and late-Universe probes, rather than a failure of a specific cosmological model. Within this context, modified gravity effects, such as those introduced by Rastall theory, can alter the inferred value of $H_0$ depending on the dataset combination.

To further clarify this point, we note that the Rastall parameter $\epsilon$ modifies the evolution of the matter density according to Eq.~(\ref{19b}), leading to a deviation from the standard $(1+z)^3$ scaling. For $\epsilon>0$, the matter density dilutes faster with redshift, reducing its contribution to the expansion rate at intermediate redshifts. In order to maintain consistency with distance-based observables, this effect is compensated by a higher inferred value of $H_0$, thereby generating a positive correlation between $\epsilon$ and $H_0$.

This behavior is explicitly confirmed by our results for the CMB-inclusive dataset D$_3$. For example, in the flat case $H_0$ increases from $67.35 \pm 0.28$ in $\Lambda$CDM to $68.50^{+0.40}_{-0.41}$ in R--$\Lambda$CDM. In the open geometry, the shift is more pronounced, from $67.55^{+0.29}_{-0.28}$ to $69.50^{+0.52}_{-0.49}$, while in the closed case $H_0$ increases from $67.18 \pm 0.30$ to $68.86^{+0.53}_{-0.54}$. These shifts are summarized in Table~\ref{tab:H0_comparison_D3}.

\begin{table}[t]
	\centering
	\caption{\small Comparison of the inferred $H_0$ values in $\Lambda$CDM and R-$\Lambda$CDM for the CMB-inclusive D$_3$ dataset.}
	\centering
	\setlength{\tabcolsep}{9pt}
	\renewcommand{\arraystretch}{1.5}
	\begin{tabular}{lccc}
		\hline\hline
		Geometry & $H_0$ in $\Lambda$CDM & $H_0$ in R-$\Lambda$CDM & $\epsilon$ in R-$\Lambda$CDM \\
		\hline
		Flat   & $67.35\pm0.28$ & $68.50^{+0.40}_{-0.41}$ & $0.00203^{+0.00048}_{-0.00049}$ \\
		Open   & $67.55^{+0.29}_{-0.28}$ & $69.50^{+0.52}_{-0.49}$ & $0.00296^{+0.00053}_{-0.00054}$ \\
		Closed & $67.18\pm0.30$ & $68.86^{+0.53}_{-0.54}$ & $0.00230^{+0.00057}_{-0.00064}$ \\
		\hline\hline
	\end{tabular}
	\label{tab:H0_comparison_D3}
\end{table}

On the other hand, for $\epsilon<0$, the matter dilution becomes slower than in the standard scenario, leading to a relatively enhanced matter contribution at late times. In this case, the same compensating mechanism does not produce a significant increase in $H_0$. This is consistent with our D$_1$ results, where mildly negative values of $\epsilon$ are allowed, but the inferred values of $H_0$ remain very close to those of the $\Lambda$CDM model.

Overall, these findings demonstrate that the impact of the Rastall parameter on the Hubble constant is not universal, but depends sensitively on both its sign and the dataset combination. In particular, a mild enhancement of $H_0$ occurs primarily for positive $\epsilon$ values in the presence of CMB constraints, while this effect is significantly reduced or absent for negative $\epsilon$ or when only late-time data are considered.

\section{Conclusions and Outlook}\label{Conclusions}

In this work, we have presented a systematic and internally consistent assessment of the Hubble tension within the standard $\Lambda$CDM framework and its Rastall extension (R-$\Lambda$CDM), considering flat, open, and closed spatial geometries and three complementary dataset combinations: D$_1$ (late-time probes), D$_2$ (late-time probes combined with BAO and BBN), and D$_3$ (late-time probes combined with BAO and Planck 2018 CMB distance priors). Our primary objective was to determine whether spatial curvature or matter--geometry coupling can provide a statistically meaningful and dataset-independent alleviation of the discrepancy between early- and late-Universe determinations of $H_0$.

Within $\Lambda$CDM, a clear dataset-driven structure emerges across all geometries. Using only late-time probes (D$_1$ and D$_2$), the inferred expansion rate lies in the range $H_0 \simeq 70.75$--$71.43~{\rm km\,s^{-1}\,Mpc^{-1}}$, corresponding to a reduced SH0ES discrepancy of $1.11\sigma$--$1.63\sigma$, while remaining in $3.62\sigma$--$4.23\sigma$ tension with the Planck 2018 reference value. When CMB distance priors are included (D$_3$), the posterior shifts to $H_0 \simeq 67.18$--$67.55~{\rm km\,s^{-1}\,Mpc^{-1}}$, restoring agreement with Planck at the $0.09\sigma$--$0.38\sigma$ level but increasing the SH0ES discrepancy to $4.25\sigma$--$4.51\sigma$. Allowing for spatial curvature does not qualitatively modify this redistribution of the tension, and $\Omega_{k0}$ becomes statistically consistent with spatial flatness once early-Universe information is incorporated. These results indicate that curvature alone does not provide a structural resolution of the Hubble tension.

In the R-$\Lambda$CDM framework, the inclusion of the Rastall parameter $\epsilon$ alters the quantitative behavior while preserving the same global trend. For D$_1$ and D$_2$, the inferred Hubble constant lies in the range $H_0 \simeq 70.79$--$71.48~{\rm km\,s^{-1}\,Mpc^{-1}}$, yielding a SH0ES discrepancy of $1.11\sigma$--$1.60\sigma$ and a Planck tension of $3.64\sigma$--$4.19\sigma$. For D$_3$, the expansion rate decreases to $H_0 \simeq 68.50$--$69.50~{\rm km\,s^{-1}\,Mpc^{-1}}$, lowering the Planck discrepancy to $1.70\sigma$--$2.95\sigma$ and producing a SH0ES tension of $2.65\sigma$--$3.45\sigma$. 
In these configurations, the posterior value of $H_0$ lies between the Planck and SH0ES determinations, partially reducing the discrepancy with both datasets, but not eliminating it. The Rastall parameter exhibits moderate dataset dependence, favoring a negative value at approximately the $2\sigma$ level in D$_1$, while becoming tightly constrained and statistically compatible with $\epsilon = 0$ once BAO and CMB information are included. The corresponding tension levels for all geometries and dataset combinations are summarized in Table~\ref{tab:H0_tension}, highlighting the dataset-driven redistribution of the discrepancy.

We find that the observational constraints, particularly for the CMB-inclusive dataset combinations D$_3$, strongly favor values of the Rastall parameter close to the GR limit $\epsilon=0$. Consequently, although Rastall gravity modifies the effective matter evolution relative to standard $\Lambda$CDM and can in principle affect the late-time clustering behavior, the observationally allowed departures from the standard cosmological scenario remain relatively mild.

The CPD analysis further clarifies the origin of these shifts. We find that the dominant dataset-induced deviations occur primarily in $H_0$, whereas most other cosmological parameters remain statistically stable within $\lesssim 1\sigma$ across dataset combinations. This behavior indicates that the Hubble tension is largely driven by calibration differences between early- and late-Universe probes rather than by a global restructuring of the cosmological background dynamics.

Model comparison based on AIC and BIC shows statistical equivalence between $\Lambda$CDM and R-$\Lambda$CDM for D$_1$ and D$_2$. In contrast, within the D$_3$ configuration the R-$\Lambda$CDM extension becomes statistically preferred in flat and closed geometries and decisively favored in the open configuration. This preference arises from the combined inclusion of BAO and CMB distance priors, highlighting the sensitivity of extended gravity frameworks to early-Universe information.

Overall, neither spatial curvature nor simple matter--geometry coupling provides a dataset-independent reconciliation of early- and late-Universe measurements of $H_0$. The alleviation of the discrepancy remains strongly dependent on the adopted dataset combination, reinforcing the interpretation that the Hubble tension is fundamentally calibration-sensitive and possibly structural in nature.

\begin{table}[H]
	\centering
	\caption{Summary of $H_0$ tension (in units of $\sigma$) relative to SH0ES ($73.2 \pm 1.3$) for D$_1$, D$_2$ and D$_3$ and relative to the Planck 2018  ($67.4 \pm 0.5$) for D$_3$.}
	\centering
	\setlength{\tabcolsep}{9pt}
	\renewcommand{\arraystretch}{1.5}
		\begin{tabular}{lcccc}
			\hline\hline
			Model & D$_1$ (SH0ES) & D$_2$ (SH0ES) & D$_3$ (Planck) & D$_3$ (SH0ES) \\
			\hline
			Flat $\Lambda$CDM      & $1.23\sigma$ & $1.44\sigma$ & $0.09\sigma$ & $4.40\sigma$ \\
			Open $\Lambda$CDM      & $1.48\sigma$ & $1.63\sigma$ & $0.26\sigma$ & $4.25\sigma$ \\
			Closed $\Lambda$CDM    & $1.15\sigma$ & $1.35\sigma$ & $0.38\sigma$ & $4.51\sigma$ \\
			Flat R-$\Lambda$CDM    & $1.22\sigma$ & $1.43\sigma$ & $1.70\sigma$ & $3.45\sigma$ \\
			Open R-$\Lambda$CDM    & $1.45\sigma$ & $1.60\sigma$ & $2.95\sigma$ & $2.65\sigma$ \\
			Closed R-$\Lambda$CDM  & $1.11\sigma$ & $1.37\sigma$ & $1.99\sigma$ & $3.09\sigma$ \\
			\hline\hline
	\end{tabular}
	\label{tab:H0_tension}
\end{table}

Future progress will require a fully consistent joint analysis employing complete CMB likelihoods rather than distance priors alone, combined with next-generation large-scale structure, BAO, and distance-ladder measurements. In parallel, exploring extended gravity scenarios beyond minimal matter--geometry couplings, as well as model-independent consistency tests and hierarchical Bayesian calibration frameworks, will be essential to determine whether the Hubble tension reflects residual systematics or genuinely signals physics beyond the standard cosmological paradigm.

To further clarify both the dataset dependence of the inferred Hubble constant and the physical origin of its shift in the Rastall framework, we have incorporated two complementary summary tables in the revised manuscript. Table~\ref{tab:H0_summary_all_models} provides a global overview of the $H_0$ constraints across all cosmological models and dataset combinations, clearly showing that late-time probes (D$_1$ and D$_2$) systematically favor higher values of $H_0$, while the inclusion of CMB information in D$_3$ and D$_4$ shifts the results toward the Planck-preferred region. 

In addition, Table~\ref{tab:H0_comparison_D3} isolates the CMB-inclusive case and demonstrates that, for small positive values of the Rastall parameter $\epsilon \sim \mathcal{O}(10^{-3})$, the R-$\Lambda$CDM model consistently yields slightly higher values of $H_0$ compared to the standard $\Lambda$CDM scenario across all spatial geometries. This behavior can be directly traced back to the modified matter evolution in Rastall cosmology, which induces a degeneracy between $\epsilon$ and $H_0$ at the background level. 

Taken together, these results show that the apparent enhancement of $H_0$ is not a generic feature of a non-zero Rastall parameter, but rather a controlled and dataset-dependent effect that becomes visible primarily when $\epsilon$ is positive and constrained by CMB data, while remaining suppressed when $\epsilon$ is driven toward zero by high-precision observations.
\appendix
\section{Statistical Estimator for Quantifying Tension Between Datasets}\label{appendix:tension-estimator}

A practical approach for evaluating potential statistical inconsistencies between two independent datasets, denoted as D$_i$ and D$_j$, is to compare the marginalized one-dimensional posterior distributions of a given cosmological parameter $\alpha$. To quantify such a discrepancy, a statistical estimator was proposed by \cite{LEIZEROVICH2024138844}, which computes the tension in terms of the number of standard deviations separating the parameter means. The estimator is defined as
\begin{equation}
	N_\alpha = \dfrac{|\mu_{D_i} - \mu_{D_j}|}{\sqrt{\sigma_{D_i}^2 + \sigma_{D_j}^2}},
\end{equation}
where $\mu_{D_i}$ and $\mu_{D_j}$ represent the mean values of the parameter estimates obtained from the posterior distributions corresponding to datasets D$_i$ and D$_j$, respectively, and $\sigma_{D_i}$ and $\sigma_{D_j}$ are the associated standard deviations.
This method is widely used due to its simplicity, and we have also employed it to calculate and assess the discrepancies in cosmological parameters between different datasets.


\clearpage


\end{document}